\documentclass[acmsmall]{acmart}

\AtBeginDocument{%
  \providecommand\BibTeX{{%
    \normalfont B\kern-0.5em{\scshape i\kern-0.25em b}\kern-0.8em\TeX}}}

\copyrightyear{2021}
\acmYear{2021}

\renewcommand{\b}[1]{\textbf{#1}}
\renewcommand{\i}[1]{\textit{#1}}

\newif\ifcomment
\commentfalse

\ifcomment
\newcommand{\hide}[1]{}

\newcommand{\added}[1]{\textcolor[rgb]{0.1, 0.56, 1}{#1}}

\newcommand{\cut}[1]{\textcolor[rgb]{0.5,0.5,0.5}{CUT: #1}}

\newcommand{\todo}[1]{\textcolor{red}{TODO: #1}}

\newcommand{\deleted}[1]{\textcolor[rgb]{0.8,0.8,0.8}{#1}}

\newcommand{\nuwan}[1]{\textcolor{blue}{NUWAN: #1}}
\newcommand{\shan}[1]{\textcolor[rgb]{0,0.8,0.4}{SHAN: #1}}
\newcommand{\zsd}[1]{\textcolor[rgb]{0.5,0.1,0.8}{ZSD: #1}}

\else
\newcommand{\hide}[1]{}
\newcommand{\added}[1]{#1}

\newcommand{\cut}[1]{}

\newcommand{\todo}[1]{}

\newcommand{\deleted}[1]{}

\newcommand{\nuwan}[1]{}
\newcommand{\shan}[1]{}
\newcommand{\zsd}[1]{}
\fi

\renewcommand{\quote}[1]{\textit{``#1''}}
\newcommand{\quoteby}[2]{\textit{``#2 (#1)''}}

\newcommand{\factor}[1]{\textbf{\textit{#1}}.}

\newcommand{\pbonf}[1]{$p_{bonf}#1$}

\newcommand{\meansd}[2]{$M = #1,\: SD = #2$} 
\newcommand{\range}[2]{$MIN = #1,\: MAX = #2$} 
\newcommand{\medianiqr}[3]{$#1 \: (#2-#3)$} 
\newcommand{\kruskalwallis}[3]{Kruskal-Wallis test, $H(#1)\:=\:#2$, #3} 
 
\newcommand{\spearmancorr}[2]{$Spearman's \: \rho = #1,\: \text{p} = #2$} 
\newcommand{\andersondarling}[2]{\i{Anderson-Darling}$ = #1,\: p = #2$} 

\newcommand{\willing}[3]{willingness $M = #1, \: SD = #2, \: n = #3$}

\newcommand{\visualduration}[5]{$#5\%, \: #1 \: (#2) \: [#3, #4]$} 
\newcommand{\visualfrequency}[2]{$#2\%, \:#1$} 
\newcommand{\participantproportion}[2]{\textit{#1 out of #2}} 
\newcommand{\participantcount}[1]{\textit{#1 participants}} 

\newcommand{\envinterrupts}[0]{\textit{external interruptions}}
\newcommand{\Envinterrupts}[0]{\textit{External interruptions}}

\newcommand{\Glance}[0]{\textit{Glance}}
\newcommand{\glance}[0]{\textit{glance}}
\newcommand{\Browse}[0]{\textit{Drift}}
\newcommand{\browse}[0]{\textit{drift}}
\newcommand{\Observe}[0]{\textit{Inspect}}
\newcommand{\observe}[0]{\textit{inspect}}

\newcommand{\Intervene}[0]{\textit{Intervene}}

\newcommand{\activebrowse}[0]{\textit{active drift}}
\newcommand{\Activebrowse}[0]{\textit{Active drift}}
\newcommand{\passivebrowse}[0]{\textit{passive drift}}
\newcommand{\Passivebrowse}[0]{\textit{Passive drift}}

\newcommand{\covertlearning}[0]{``covert learning''}

\newcommand{\signal}[0]{\textit{signal}}
\newcommand{\Signal}[0]{\textit{Signal}}
\newcommand{\signalforperception}[0]{\textit{background signal}}
\newcommand{\signalforaction}[0]{\textit{trigger signal}}
\newcommand{\Signalforperception}[0]{\textit{Background signal}}
\newcommand{\Signalforaction}[0]{\textit{Trigger signal}}

\newcommand{\phoneprobe}[0]{\textit{phone probe}}
\newcommand{\glassprobe}[0]{\textit{OHMD probe}}

\newcommand{\studyone}[0]{\textit{study 1}}
\newcommand{\Studyone}[0]{\textit{Study 1}}
\newcommand{\studytwo}[0]{\textit{study 2}}
\newcommand{\Studytwo}[0]{\textit{Study 2}}

\newcommand{\aprox}[1]{$\approx$}

\newcommand{\plus}[0]{$(+)$}
\newcommand{\minus}[0]{$(-)$}
\newcommand{\neutral}[0]{$(\bullet)$}

\usepackage{subcaption}
\usepackage{enumerate}
\usepackage[shortlabels]{enumitem}
\usepackage{multirow}
\usepackage[normalem]{ulem}
\useunder{\uline}{\ul}{}

\usepackage{arydshln}

\begin{document}

\title{Visual Behaviors and Mobile Information Acquisition}

\author{Nuwan Janaka}
\email{nuwanj@comp.nus.edu.sg}
\orcid{0000-0003-2983-6808}

\affiliation{%
  \institution{National University of Singapore}
  \department{NUS-HCI Lab, School of Computing}
  \country{Singapore}
}

\author{Xinke Wu}
\authornote{Authors contributed equally to this research.}
\email{xinke.wxk@antgroup.com}

\affiliation{%
  \institution{National University of Singapore}
  \department{NUS-HCI Lab, School of Computing}
  \country{Singapore}
}

\affiliation{%
  \institution{Alibaba}
  \streetaddress{Z Space, No. 556 Xixi Road}
  \city{Hangzhou}
  \country{China}
  \postcode{310000}
}

\author{Shan Zhang}
\authornotemark[1]
\email{shan_zhang@u.nus.edu}
\orcid{0000-0003-1030-537X}

\affiliation{%
  \institution{National University of Singapore}
  \department{NUS-HCI Lab, School of Computing}
  \country{Singapore}
}

\author{Shengdong Zhao}
\email{zhaosd@comp.nus.edu.sg}
\orcid{0000-0001-7971-3107}

\affiliation{%
  \institution{National University of Singapore}
  \department{NUS-HCI Lab, School of Computing}
  \country{Singapore}
}

\author{Petr Slovak}
\email{petr.slovak@kcl.ac.uk}
\orcid{0000-0001-8458-7715}

\affiliation{%
  \institution{King's College London}
  \department{Department of Informatics}
  \streetaddress{Bush House, 30 Aldwych}
  \city{London}
  \country{United Kingdom}
  \postcode{WC2B 4BG}
}

\renewcommand{\shortauthors}{Janaka, et al.}


\begin{abstract}
It is common for people to engage in information acquisition tasks while on the move. To understand how users' visual behaviors influence microlearning, a form of mobile information acquisition, we conducted a shadowing study with 8 participants and identified three common visual behaviors: \glance{}, \observe{}, and \browse{}. We found that \browse{} best supports mobile information acquisition. We also identified four user-related factors that can influence the utilization of mobile information acquisition opportunities: situational awareness, switching costs, ongoing cognitive processes, and awareness of opportunities. We further examined how these user-related factors interplay with device-related factors through a technology probe with 20 participants using mobile phones and optical head-mounted displays (OHMDs). Results indicate that different device platforms significantly influence how mobile information acquisition opportunities are used: OHMDs can better support mobile information acquisition when visual attention is fragmented. OHMDs facilitate shorter visual switch-times between the task and surroundings, which reduces the mental barrier of task transition. Mobile phones, on the other hand, provide a more focused experience in more stable surroundings. Based on these findings, we discuss trade-offs and design implications for supporting information acquisition tasks on the move.

\end{abstract}



\maketitle

\section{Introduction}

Think about the last time you commuted: how many times did you shift your attention away from your phone to navigate a busy street or look at a signboard? 

Visual attention is a critical resource for processing visual information on the move. This process includes filtering information that one receives, then selectively processing the content \cite{carrasco_visual_2011, connor2004visual, chun_visual_2005}. It is more difficult to perform information processing tasks with computing devices while on the move than in stationary settings as higher levels of attention are required for both the mobility task (e.g., walking) and mobile Human-Computer Interaction (HCI) task (e.g., reading an email) \cite{wobbrock_situationally_2019, oulasvirta2005interaction}. Mobility tasks require users to focus on their surroundings in an effort to maintain situational safety as well as react to social or personal need-based cues, while mobile HCI tasks require users to maintain visual attention on their device \cite{spink2008multitasking, wobbrock_situationally_2019, stavrinos_distracted_2011}. Consequently, visual attention fragmentation occurs more frequently when users are on the move \cite{oulasvirta2005interaction}. 

The topic of visual attention allocation (or, more broadly, visual behaviors \cite{henderson_human_2003, ellsworth_visual_1972}) is an important area of research that has been extensively studied. Yet, previous investigations have mostly focused on visual behaviors associated with stationary settings \cite{rayner_eye_1998, carrasco_visual_2011, land_eye_2006}. Relatively few studies have concentrated on visual behaviors in mobile contexts \cite{steil2018forecasting, oulasvirta2005interaction}, and there has been a lack of categorization around visual behaviors in this context. In this study, we aim to deepen our understanding of visual behavior patterns on the move, as well as precise and in-situ mobile interaction designs to support mobile information acquisition tasks. While previous studies have investigated the fragmented nature of visual behavior in mobile settings \cite{oulasvirta2005interaction}, we aim to examine the different aspects of fragmentation, formally classify them, and investigate their influence on the effectiveness of information acquisition in mobile contexts.

We contextualized our investigation in a specific mobile scenario (commuting) and information acquisition task (microlearning vocabulary). Commuting was selected as it is a typical mobile situation in everyday life \cite{knupfer2018elements}. It is naturally accompanied by complex and dynamic external distractions (e.g., people moving, sudden noises), which influence on-the-move information processing \cite{shaw_its_2019, castellano_sensorimotor_2016}. Since visual attention is often fragmented during on-the-move situations \cite{oulasvirta2005interaction}, we avoided using long/complex information acquisition tasks as they can hinder the resumption of ongoing (mobility) tasks \cite{monk_effect_2008, cai_waitsuite_2017}. Instead, we focused on information acquisition tasks that consist of smaller tasks with fewer dependencies. Microlearning is well-suited for this purpose \cite{dingler_language_2017} as the microlearning technique divides complex learning tasks into small ``bite-sized'' sessions and integrates them into daily activities \cite{gassler_integrated_2004}.

We identified three distinct visual behaviors resulting from attention fragmentation while commuting through a shadowing study: \Observe{}, \Browse{} and \Glance{}, based on the dimensions of \i{Purpose},  \i{Duration}, and \i{Perceived Visual Attention Intensity}. 
We studied how these patterns offer different opportunities for mobile information acquisition tasks on the move, specifically for microlearning during the commuting scenario. We found that \browse{} presents the most suitable opportunity for mobile information acquisition. However, the utilization of visual behaviors largely depends on dynamic interruptions from the environment (\envinterrupts{}).

Mobile information acquisition and interactions are tied to specific devices. Mobile phones are the most common platform at present, but existing research shows that on-the-move information acquisition with mobile phones leads to fatigue and reduced learning gains \cite{zhao_stationary_2018, castellano_sensorimotor_2016, khan_designing_2020, wobbrock_situationally_2019}. Optical see-through Head Mounted Displays (OST HMDs, OHMDs) or smart glasses are an emerging mobile interaction platform that have been shown to minimize the issue of split attention. This platform can provide peripheral information to users, reducing interferences between the surrounding environment and on-the-move mobile interactions \cite{lucero_notifeye_2014, maples2008effects, spitzer_distance_2018, ishiguro_peripheral_2011}.

For more insight into visual interaction designs during on-the-move situations, we investigated how different devices (mobile phones and OHMDs) utilize \browse{s} differently for second-language microlearning during commute through a technological probe. Results revealed that OHMDs enabled a better balance between information acquisition tasks and situational awareness and allowed commuters to utilize shorter \browse{s} amid of frequent \glance{s} for information acquisition. 
On the other hand, mobile phones provided a more focused experience for mobile information acquisition when the surroundings were more stable and had fewer \envinterrupts{}.

Based on these findings, we discuss the trade-offs and design implications for supporting mobile information acquisition on the move, especially for microlearning on the commute, and propose a system that can utilize opportunistic visual behaviors in a more general setting.

\noindent{}Our contribution is twofold: 

1) We establish commuters' visual behaviors with respect to three categories (\glance{}, \observe{}, and \browse{}) and identify their effects on mobile information acquisition (i.e., microlearning) opportunities. We discuss design implications for better supporting mobile HCI tasks.

2) We empirically evaluate the \i{receptivity} for mobile information acquisition on both mobile phones and OHMDs during a dynamic mobile context (e.g., commute) and identify the trade-offs of using both platforms. In doing so, we better understand how device platforms affect mobile information acquisition and missing interactions.

\section{Related Work}
\label{sec:related_work}

Our work relates to three broad areas. 

\subsection{Multitasking, attention fragmentation, and visual behaviors}

There are multiple theories and frameworks regarding attention management and allocation across various tasks, such as Kahneman's resource theory \cite{kahneman1973attention}, Wicken's multiple resource theory \cite{wickens_processing_1991}, and the Resource Completion Framework \cite{oulasvirta2005interaction}. In each of these theories, attention is modeled as a finite (or elastic) resource in which multitasking can be cognitively, perceptually, physiologically, and socially costly \cite{wobbrock_situationally_2019, oulasvirta2005interaction}. In multitasking scenarios, attention is simultaneously shared across different tasks \cite{wickens_processing_1991, oulasvirta2005interaction}, which leads to attention fragmentation \cite{oulasvirta2005interaction}.
Despite the attention costs involved in multitasking, people continue to multitask with their mobile devices on the move \cite{wang2012myth, paridon2010multitasking}, suggesting that there is a strong demand for information acquisition on the move. We seek to investigate how visual attention is allocated as an essential step to supporting this demand. 


To understand various aspects of attention in real-world mobile HCI tasks in an effort to design interfaces for limited attention spans, HCI researchers have investigated attention allocation on the device or environment based on task levels \cite{bace_quantification_2020, steil2018forecasting, oulasvirta2005interaction}. For example, establishing that users' attention span is 4 to 8 seconds on mobile devices \cite{oulasvirta2005interaction} has helped designers to size information chunks accordingly such that information can be effectively consumed in a short duration or glance. 

However, the current understanding of the topic (e.g., duration of attention fragments) has proved insufficient to guide detailed designs on when and how to present information to users in mobile scenarios \cite{steil2018forecasting, bace_quantification_2020}. Our study investigates visual behavior patterns in terms of duration, purpose, and intensity when users are on the move. With that, we develop a better understanding of various visual behaviors and their opportunities for effective information acquisition.


\subsection{Mobile HCI, information acquisition, and microlearning}

It is common for people to engage in mobile HCI tasks while on the move and the majority of these tasks are for the purpose of information acquisition (e.g., reading articles, checking social network updates, watching videos) \cite{guo_smart_2015, russell_what_2011, oulasvirta2005interaction}.

Microlearning is one such information acquisition technique that divides complex learning tasks into small and quick learning interactions distributed across time \cite{hutchison_context_sensitive_2007, gassler_integrated_2004}.
Microlearning is commonly applied to language learning as it is relatively easy to break down language learning tasks such as vocabulary learning into smaller ones \cite{cates_mobilearn_2017, gassler_integrated_2004, edge_micromandarin_2011}. Various methods of improving mobile language microlearning such as adaptations to individual learners \cite{edge_memreflex_2012}, spaced repetitions \cite{edge_micromandarin_2011, webb2007effects}, multi-modal presentations of content \cite{cates_mobilearn_2017}, and contextual encoding \cite{dearman_evaluating_2012, edge_micromandarin_2011, trusty_augmenting_2011, hutchison_context_sensitive_2007} have been explored in previous studies. 

Investigators also have researched on \textit{when} to present microlearning content, in which studies focused on identifying ubiquitous micro-moments during daily life when users are less engaged with their mobile devices \cite{cai_waitsuite_2017, cai_wait_learning_2015, gassler_integrated_2004, ren_pull_refresh_2015}. Some research has honed in on the internal factors (e.g., boredom \cite{dingler_language_2017}) that influence microlearning opportunities as a means of identifying suitable timing independent of tasks. 


However, how visual attention influences microlearning opportunities remains an underexplored area. This work further undertakes this area of research by investigating the interplay between visual attention with opportunistic moments for microlearning, its implications so that this knowledge can be applied more broadly to general information acquisition tasks. 

\subsection{Influence of platforms on mobile HCI tasks}

Mobile phones are the most commonly used platform for information acquisition tasks such as language microlearning \cite{edge_micromandarin_2011, cates_mobilearn_2017, webb2007effects, cai_wait_learning_2015, ren_pull_refresh_2015, dearman_evaluating_2012}. With the advancement in mobile technologies, users can access information anytime and anywhere without physical and social boundaries \cite{perry_dealing_2001, naismith_literature_2004}. However, mobility presents interactional challenges due to constant situational and contextual changes in the user's environment, and it is difficult for users to sustain such high levels of attention on HCI tasks \cite{sharples_mobile_2009, perry_dealing_2001}.

OHMDs, on the other hand, have emerged as a promising platform to support multitasking as users can maintain direct visual contact with their physical surroundings while performing mobile HCI tasks displayed on screen \cite{orlosky_managing_2014, lucero_notifeye_2014, rauschnabel_augmented_2015}. 

Both mobile phones and OHMDs can support mobile information acquisition tasks such as microlearning due to their portability, ease of access, and support of contextual encoding \cite{edge_micromandarin_2011}, though there are trade-offs to each platform. For instance, mobile phone users do not have as much access to their peripheral vision when interacting heads-down with their phones \cite{maples2008effects}; thus, any information displayed on the mobile screen is potential `distraction' for their mobility tasks \cite{elder_technical_2015}. In contrast, OHMD users have better visual access to their environment, given that their visual headset is worn and their display remains perpetually in view \cite{luyten_hidden_2016}. Still, focusing on learning content on a see-through display can be more challenging than on a mobile screen, as the background can change.

This research intends to investigate the advantages and disadvantages of information acquisition opportunities offered by the two platforms.

\section{Overall Study Design}

Multitasking on the daily commute is common and habitual for most \cite{shaw_its_2019}. As previously highlighted, a key issue with such behaviors is that it requires visual attention to split between navigational and mobile HCI tasks \cite{oulasvirta2005interaction}. We unpack our understanding of this problem by first conducting an observational shadowing study (\studyone{}), followed by a technology probe (\studytwo{}) on mobile phones and OHMDs, comparing on-the-move microlearning between the two platforms.


\section{Study 1: Understanding visual behaviors during commuting}

In this first study, we focus on exploring the following research questions.
\vspace{1mm}

\noindent{}\b{RQ1:} What are the typical on-the-move visual behaviors when factoring for purpose, duration, and intensity? What are the observable characteristics of visual behaviors?

Previous work has shown that a person's attention shifts based on the purpose, duration, and perceived intensity of focus change \cite{mccallum_attention_2021}. We adopt these categories as dimensions in our investigation of visual behaviors.  

Since purpose and perceived intensity are subjective in nature and cannot be directly measured, we focused on head movements (often used with gaze estimation) \cite{valenti_combining_2012} as the observable trait associated with purpose and intensity. 
\vspace{1mm}

\noindent{}\b{RQ2:} How can visual behaviors support information acquisition tasks such as microlearning?
\vspace{1mm}

There are many opportunities to divert attention to information acquisition tasks when a person is on the move.  We explore the visual behaviors that support such information tasks and the effect of behavior switching.

\subsection{Method}
We first carried out a shadowing study (with video recording) to investigate real-life individual and social interactions in-situ and the effects of the surroundings \cite{quinlan_conspicuous_2008, asan_using_2014}. 

We also used contextual inquiries \cite{lazar_2010_research_method}[Ch~8] to understand how participants can potentially use visual attention and how receptive they are towards diverting attention to microlearning. Following Isaacs et al. \cite{isaacs_mobile_2009}, we operationalized receptivity as the willingness to engage with microlearning. To identify the \i{instantaneous} and \i{in-situ receptivity} in dynamic commuting scenarios, we conducted contextual inquiries in a fashion similar to event-contingent experience sampling with verbal responses \cite{napa_scollon_experience_2009, consolvo_using_2003}. 
To reduce interruptions to natural behaviors, each inquiry was at least 15-minutes apart and had a maximum duration of 3 minutes. 
Each participant was shadowed 2-3 times within the same day to increase the generalizability of the results.

\subsubsection{Apparatus}
\label{sec:study1_aparatus}
\begin{figure}[h]
  \centering
  \includegraphics[width=\linewidth]{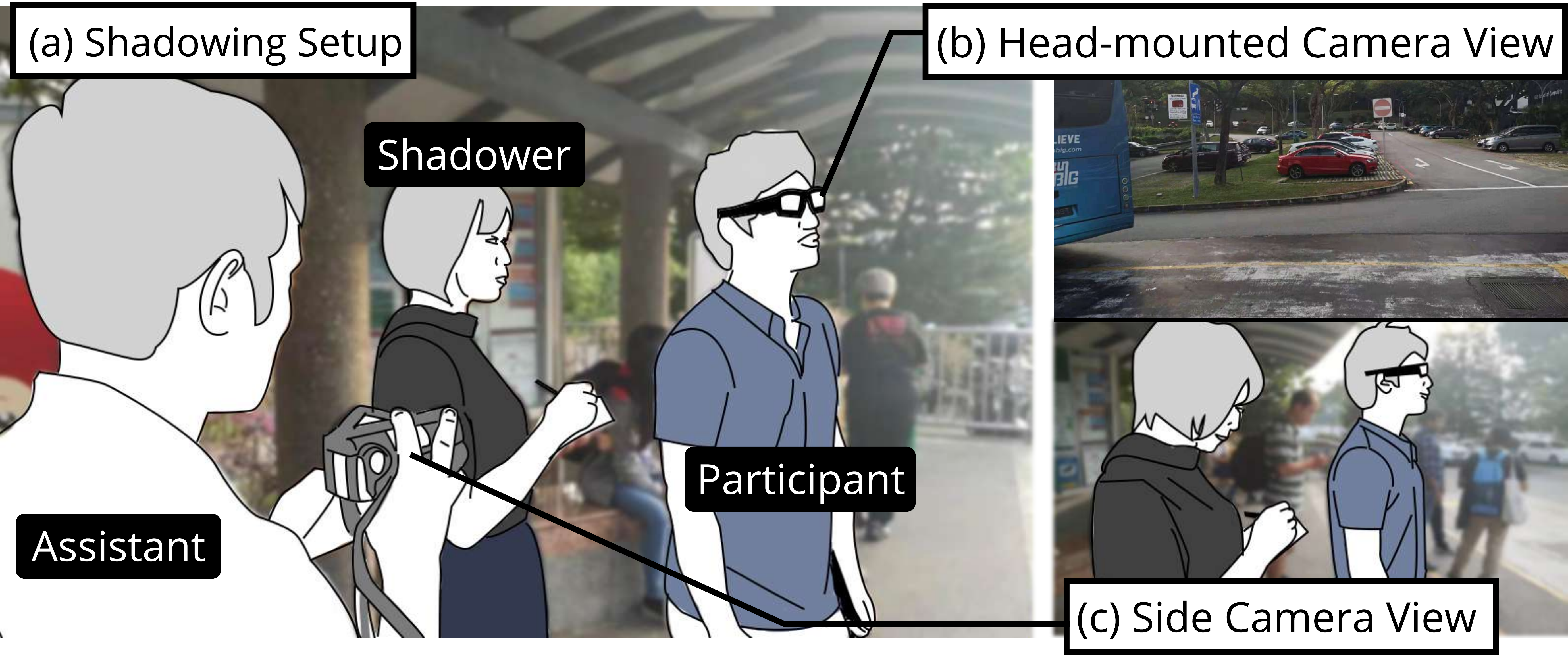}	  
  \caption{Shadowing configuration (a) The shadower observes the participant who is wearing a head-mounted camera and takes notes, while the assistant records the side view of the participant (b) Head-mounted camera view from the head-mounted camera (c) Side camera view recorded by the assistant}
  \Description{Shadowing setup and combined camera view}
  \label{fig:study1_setup}	  
\end{figure}

As our focus is on task-level visual behaviors instead of micro-level eye movements, we used video recordings that were similar to those used by Oulasvirta et al. \cite{oulasvirta2005interaction}, but only with a head-mounted camera (weight $\approx$ 24g) and side camera, as shown in Fig~\ref{fig:study1_setup}.
The head-mounted camera was for identifying the focus of users' attention, while the side view camera was for identifying attention switches (with spatial changes) during task engagement.

\subsubsection{Participants}
\label{sec:study1_participants}
Since we targeted populations with mobile information needs, we selected participants who were already personally motivated to perform information acquisition tasks on the move.  Eight participants [P1-P8] (4 females, age \meansd{23.5}{2.8}) were selected based on their language learning experiences and commuting profiles (Table~\ref{table:study1_participants}). All participants were formally educated in English, had been on two or more types of commute (metro/bus/walking), and had experience using mobile vocabulary learning apps, albeit not in all commuting scenarios. 

We compensated each participant with $\approx$ USD 7.40/h for their time in \b{both} studies and received their informed consent before conducting the studies.

\begin{table*}
\caption{Participants' demographic information and shadowing settings in \studyone{}}
\label{table:study1_participants}
\begin{tabular}{p{0.1\textwidth} p{0.23\textwidth} p{0.32\textwidth} p{0.3\textwidth} }
\toprule
\b{Participant} & \b{Gender, Age, Native language, Profession/Major} & \b{Language learning usage} & \b{Shadowing duration, Medium, Commuting purpose, Time of the day}  \\ \midrule
P1 & 
Male, 23, English, Undergraduate (Real Estate) & 
Use \i{Jisho} to learn Japanese & 
180 min, Metro/Bus/Walking, To \& from School/Dining, Morning/Evening \\ \midrule 

P2 & 
Female, 23, English, Undergraduate (Industrial and System Engineering) & 
Use \i{AnkiDroid} to learn Japanese. Do not use mobile on the bus/during walking due to motion sickness & 
220 min; Metro/Bus/Walking, To \& from School/Dining/Exercise, Morning/Noon/Evening \\ \midrule

P3 & 
Male, 22, English, Undergraduate (Linguistics)  &
Used \i{Memrise} to learn Spanish & 
150 min, Bus/Walking, To \& from School/Dining/Library, Noon/Evening \\ \midrule

P4 & 
Female, 22, Mandarin, Master Student (Finance) & 
Used \i{Shanbay} to practice English & 
140 min, Bus/Walking, To \& from School/Dining, Morning/Evening \\ \midrule

P5 & 
Male, 28, Sinhale, Software Engineer (Electronics)  & 
Used \i{Magoosh GRE} to practice English. Do not use mobile during the bus due to motion sickness & 
200 min, Bus/Metro/Walking, To \& from Work/Dining/Shopping, Morning/Noon/Evening \\ \midrule

P6 & 
Female, 21, English, Undergraduate (Environmental Engineering) & 
Use \i{TenguGo Hangul} to learn Korean. Do not use mobile while walking & 
230 min, Metro/Bus/Walking, To \& from School/Shopping, Noon/Evening \\ \midrule

P7 & 
Male, 28, Spanish, PhD Student (Design \& Environment) & 
Use \i{Duolingo} to learn Mandarin & 
200 min, Metro/Bus/Walking, To \& from School/Shopping, Morning/Afternoon 
\\\midrule 

P8 & 
Female, 21, English, Undergraduate (Economics) & 
Use \i{Duolingo} to learn French & 
240 min, Metro/Bus/Walking, To \& from School/Dining/Shopping, Morning/Evening 
\\ \bottomrule
\end{tabular}
\end{table*}

\subsection{Procedure}
\label{sec:study1_procedure}

Each participant was followed by a shadower and an assistant for 2-4 hours a day in different commuting scenarios, as shown in Fig~\ref{fig:study1_setup} (a) and \b{Table~\ref{table:study1_participants} (column 4)}. After participants were briefed and asked for consent, they put on a head-mounted camera, which recorded a view of their vision field (Fig~\ref{fig:study1_setup} (b)). 

During the study, the shadower took observational notes, and the assistant recorded the side view of the participant, especially his/her head and hand movements (Fig~\ref{fig:study1_setup} (a), (c)).
When participants' behavior or context changed, the shadower noted down the time, changes in context, as well as their attention focus. Participants were asked a set of questions, including their receptivity (i.e., willingness) to microlearning (5-point Likert scale, 1 = Very Low, 5 = Very High), factors that affected their willingness to microlearning, and the perceived amount of visual attention they paid to the primary task (5-point Likert scale, 1 = Very Low, 5 = Very High).  
The monitored contexts and details of contextual inquiries are in Appendix~\ref{appendix:study1_contexts} and Appendix~\ref{appendix:study1_contextul_inquiry}.

At the end of the shadowing, the shadower carried out a 30-40 minute semi-structured interview and asked participants about the reasons behind the changes in their visual behaviors, how that or other factors influenced their receptivity to microlearning (refer to Appendix~\ref{appendix:study1_interview} for interview topics). The entire interview was audio-recorded and later transcribed. Whenever required, such as when participants needed help to recall details, the assistant played them the relevant parts of the video recording.

\subsection{Data Analysis}
\label{sec:study1_data_analysis}
We conducted a mixed analysis (mainly qualitative) by triangulating four data sources: contextual inquiry notes from 112 inquiry sessions, interview transcriptions of 8 sessions, observation notes, and video recordings from approximately 26 hours of shadowing (see Appendix~\ref{appendix:study1:data_triangulation} for details). Text data (i.e., contextual inquiry notes, observation notes, and interview transcriptions) contained the perceived behaviors, reasons, and contextual information. Video data (i.e., head-mounted camera view time-synced with a side view) consisted of head/eye movement data, attention focus, and timing information. Since the two forms of data showed different but complementary dimensions of visual behaviors, we used two different coding schemes initially and combined them with the themes later.

Using the \textit{QDA Miner} software package, two researchers (co-authors and shadowers) independently performed open coding \cite{corbin_grounded_1990, corbin2014basics} on \b{two} participants' text data and video recordings. 

The two researchers then discussed and developed initial coding schemes, one for text data and another for video data. Then, they independently reanalyzed the same participants' data, resolved any disagreements, and refined the coding by discussing and re-watching video recordings. After analyzing another participant's data independently, the two coders reached 93\% agreement on text data, but the video coding timing varied by 0-13s. This discrepancy happened because the purpose of visual attention was only captured during contextual inquiries but not in all video data. 

The researchers used the resulting codes to independently analyze the remaining participants' data to develop themes and patterns by grouping codes. Two researchers watched the video recordings to reach an agreement whenever there was a discrepancy between the subsequent codes and themes. The data with discrepancies were reanalyzed using the codes that were agreed upon.

Videos were coded into datasheets by pausing and replaying them based on the agreed codes: time, posture, location, familiarity, crowd levels, and visual behavior (with an accuracy of one second - refer to Appendix~\ref{appendix:video_coding} for sample data). Observation and contextual inquiry notes were merged with the video coding based on the respective event time for (quantitative) descriptive analysis. Any time discrepancies were resolved using the video data.

\section{Findings: Study 1}
\label{study1_results}
We present our findings based on the research questions and themes that emerged from the data analysis.

\subsection{What are the typical on-the-move visual behaviors?}

\subsubsection{Visual behavior patterns}
\label{sec:visual_behavior_patterns}
\begin{figure*}[h]
  \includegraphics[width=\linewidth]{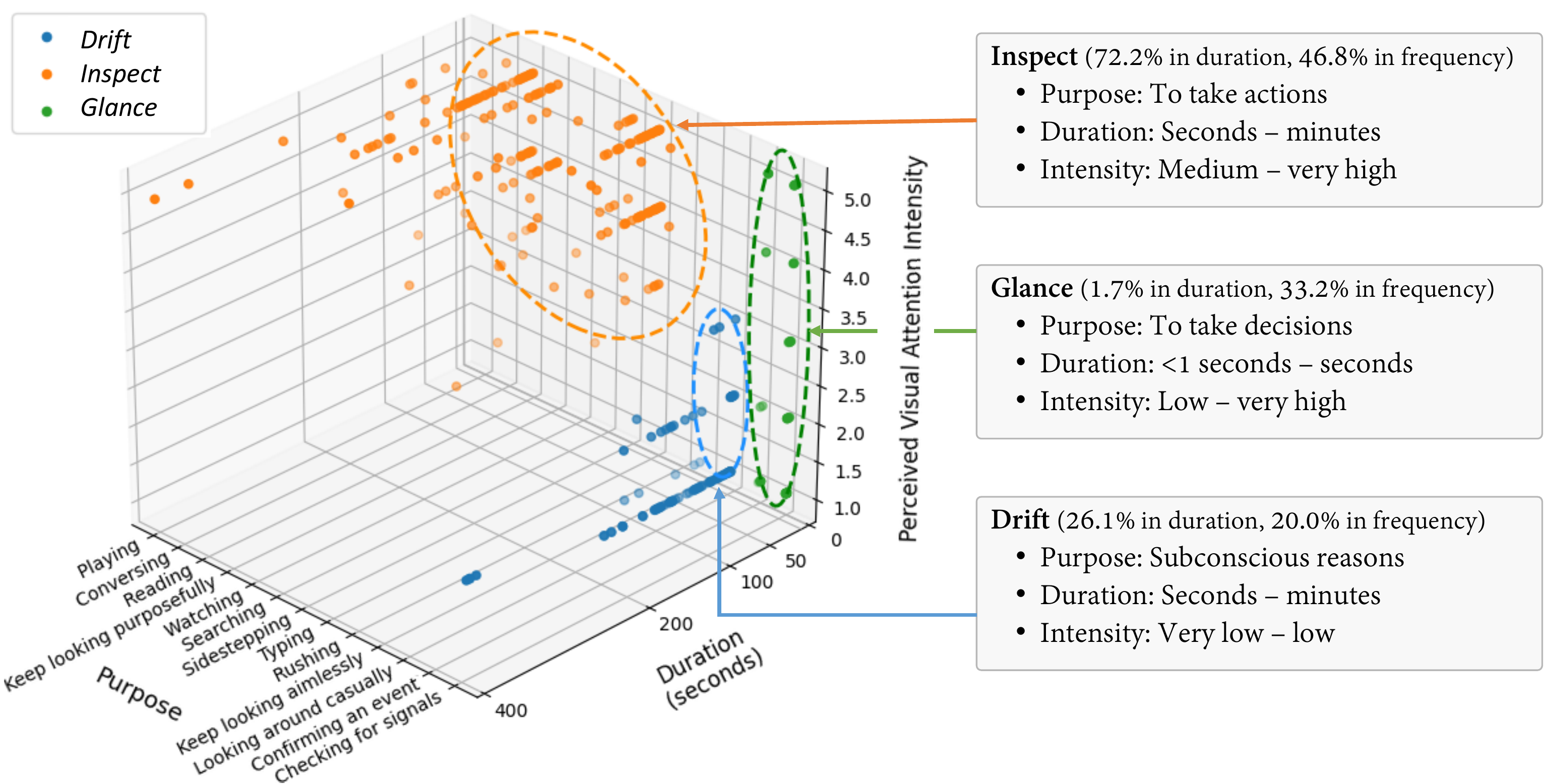}
  \caption{3D scatter plot of the three visual behavior patterns: \glance{}, \observe{}, and \browse{}, clustered by purpose, duration of the behavior, and perceived visual intensity. 
  For example, checking the arrival of a bus, confirming bus numbers, or checking for notification type are \glance{s}. Playing a mobile game, looking for someone, or watching the outside scenery intentionally are \observe{s}. Visually scanning the surroundings without intention or staring at something without purpose are examples of \browse{s}. 
  We further classified the purpose for engaging in each behavior pattern into different subcategories. Duration (seconds) is displayed on a log scale. Perceived visual attention intensity is represented as a 5-point Likert scale (1 = Very Low, 5 = Very High). See Appendix~\ref{appendix:visual_behavior_distributions} for duration distributions. NOTE: The visual patterns shown here are non-exhaustive as they only cover instances of inquiry.}
  \Description{3D scatter plot of the visual behavior patterns.}
  \label{fig:visual_patterns}	  
\end{figure*}

To analyze the visual behavior patterns in the contexts of mobile human-computer interactions, we describe the visual behaviors at the task level based on their characteristics, i.e., purpose, duration, and intensity. 
We identified three distinct patterns from the mixed analysis: \glance{}, \observe{}, and \browse{}, as illustrated in \textbf{Fig~\ref{fig:visual_patterns}}. For each pattern, we distinguish between two main reasons for engagement: decision-making and action-taking. 

\b{\Glance{}}: Commuters utilize \glance{s} for the purpose of decision-making, such as to check if action is required. \Glance{s} are less than a second to several seconds in duration since decision-making does not require much time. Furthermore, the intensity of the visual attention required for \glance{s} varies; relatively low-intensity attention focus is required when users can return to their primary task at hand without taking any post-action, such as when a commuter looks up to check the bus number, only to resume his previous task when he finds that it is not the correct bus. On the other hand, greater intensity in attention focus is required when a post-action is necessary, such as when this same commuter \glance{s} up to check the bus number and prepares to board it when he finds that it is indeed the right bus that has arrived.
During a \glance{}, the commuter's gaze and head orientation change quickly within a few seconds and in a highly alert state\footnote{the state of being attentive and prepared to react \cite{apa_alertness_2019}}. 
Checking the arrival of a bus or sudden notifications on the phone are thus examples of \glance{s}.

\b{\Observe{}}:  \Observe{} occurs when people need to keep their gaze on an object related to their task at hand or engage in action. It can last for several seconds to minutes, depending on task duration. Texting, reading, and chatting are ubiquitous activities that fall under the category of \observe{ion}. \Observe{ion} requires continuous monitoring and thus generally requires higher visual attention intensity.
During \observe{s}, the commuter's gaze and head orientation change slowly or are kept static with high alertness.
Intentionally looking for someone, reading notices, and watching the outside scenery are instances of \observe{s}.

\b{\Browse{}}: \Browse{} is a natural behavior to reduce physical fatigue, refresh the commuter's mind. Subconscious activities such as pondering, recalling past events, or daydreaming fall under this category. People usually \browse{} when they are not visually engaged with any task, such as when a commuter unintentionally looks around or stare at something to rest his mind after having read an article for some time. \Browse{} can last from seconds to minutes, depending on the duration of the interruption, or until \browse{} is switched to \glance{} or \observe{}. \Browse{} requires low levels of visual attention since it is not a full engagement of a task. 
During \browse{}, the commuter's gaze and head orientation change slowly or remain static with low levels of alertness.

\subsubsection{Visual behavior transitions}
\label{subsec:visual_behavior_transistions}

\begin{figure*}[h]
  \centering
  \includegraphics[width=\linewidth]{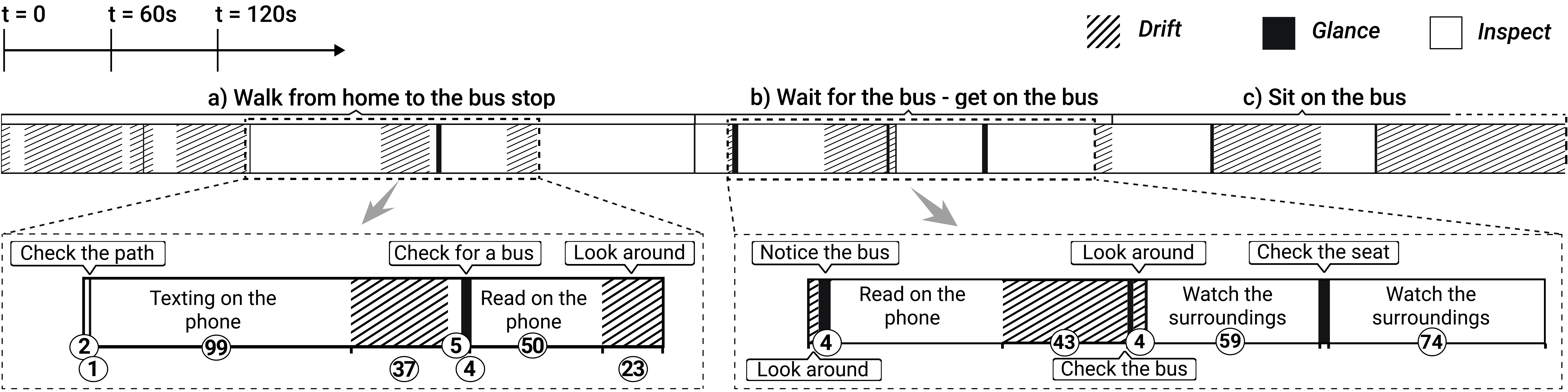}
  \caption{A participant's (P1) visual behaviors during commuting from home to school illustrate how visual behaviors fluidly transition. The numbers inside circles represent the duration for each pattern in seconds. a) Walking from home to the bus stop (time taken: 8min 49s), b) Waiting to get on the bus (time taken: 4min 50s), and c) Commuting on the bus.  Refer to Appendix \ref{appendix:video_coding} for further details.}
  \Description{Visual behavior of an individual during the commute from home to school.}
  \label{fig:commute_pattern}	  
\end{figure*}

As illustrated in Fig~\ref{fig:commute_pattern}, participants fluidly switch between patterns while performing different tasks on the move. They may occasionally \glance{} at their surroundings while \browse{ing} and \observe{ing} (e.g., Fig~\ref{fig:commute_pattern} (1) `check the path' (\glance{}) or `check for a bus' (\glance{}) during `texting/reading on the phone' (\observe{})). These \glance{s} can interrupt the ongoing tasks and divert their visual attention to new tasks/actions, such as when a commuter stops texting on his phone (\observe{}) when he notices (\glance{}) that someone is walking towards him in order to avoid collision (\observe{}).

\begin{figure}[h]
  \centering
  \includegraphics[width=\linewidth]{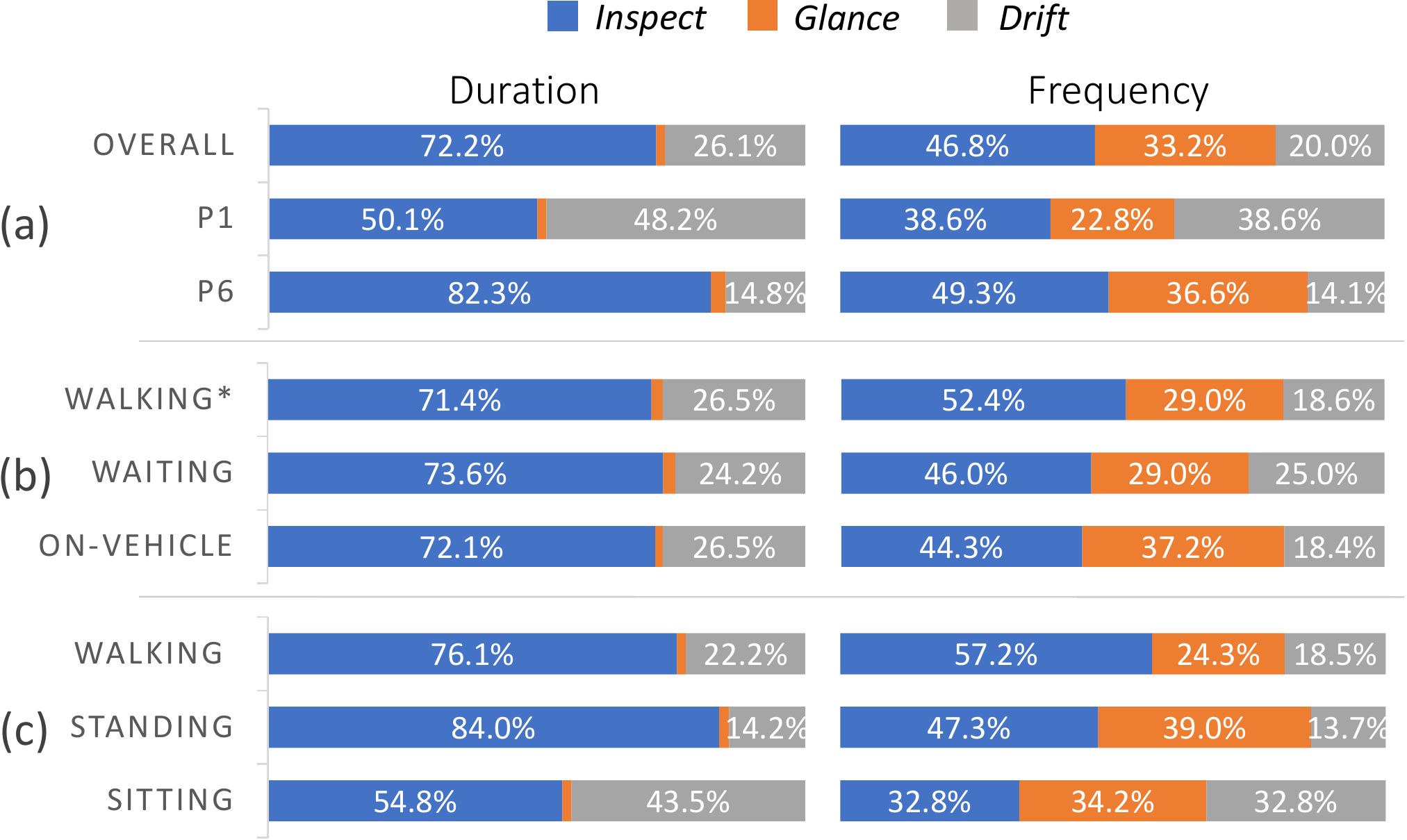}
  \caption{Overall variation in visual behaviors across duration and frequency, based on (a) individual behaviors, (b) commuting stage (walking for commuting, waiting for commuting, and on-vehicle commuting), and (c) posture. $^*$ Includes scenarios such as traveling on an escalator/lift which does not involve walking and does not belong to other commuting stages. Refer to Appendix~\ref{appendix:behavior_frequency} for further details. }
  \Description{Overall Visual behavior variation across duration and frequency.}
  \label{fig:behavior_frequency}
\end{figure}

The duration and frequency of visual patterns and transitions were dependent on contextual factors \cite{isaacs_mobile_2009} such as location, time, crowdedness, and personal habits like mobile phone usage. Two participants who showed the highest (P1, 48.2\%) and lowest (P6, 14.8\%) overall duration for \browse{s} exemplify this dependency in Fig~\ref{fig:behavior_frequency}. 

Furthermore, \glance{s} and \observe{s} were affected by mobile HCI tasks. 23.8\% in duration (17.5\% in frequency) of \glance{s} and 71.8\% of duration (65.8\% in frequency) of \observe{s} were attributed to mobile HCI tasks that occurred during commuting (e.g., \glance{s}: checking smartwatch, \observe{s}: reading/texting/watching on the phone or tapping travel card). The duration of visual patterns lasted between 1 second (\glance{}: when checking the phone/smartwatch for new notifications/time while walking) and 695 seconds (\observe{}: when playing a mobile game while standing on the metro), depending on the task and context. As expected, greater engagement with mobile HCI tasks reduced the overall duration of \browse{s} and increased the overall duration of \observe{s}.

\subsection{How can visual behaviors support information processing tasks such as microlearning?}

\subsubsection{Visual behaviors and receptivity to microlearning}
\label{subsec:behavior_and_receptivity}

Our findings suggest a significant connection between visual behavior patterns and receptivity towards microlearning. 
Comparing the receptivity ratings between patterns, we found a significant effect of \i{pattern} (\kruskalwallis{2}{198.4}{p $<$ 0.001}) and Dunn's post-hoc comparison showed significant difference (\pbonf{$<$0.001}) between each pair from high to low receptivity: \browse{} $>$ \observe{} $>$ \glance{}. 

\factor{\Browse{}} 
Generally speaking, \browse{s} presented the most opportunistic moments for microlearning since they had the highest receptivity (\willing{3.8}{0.33}{132}, \i{n} represents the number of patterns related to contextual inquiries). We differentiate between two types of \browse{}: \passivebrowse{} and \activebrowse{}. \Passivebrowse{} indicates when the participant has nothing to do, such as boredom, e.g., \quoteby{P7}{[sitting on an empty bus,] I am doing nothing now [observed \browse{}], so I think it's good to learn words}. \Activebrowse{} indicates relaxation and ponderation, e.g., \quoteby{P4}{I was thinking something important at that time [during observed \browse{}]. I don't think it's a good time [for microlearning].}. Interestingly, the observed ratio between \passivebrowse{} and \activebrowse{} was 126:6 = 95\%, indicating that users were willing to engage in more meaningful activities if given a choice for the majority of \browse{s}. Cognitive load monitoring \cite{haapalainen_psycho_physiological_2010} may help us distinguish between \passivebrowse{} and \activebrowse{} since the latter involves non-visual cognitive tasks such as pondering.

Our analysis also shows a positive, medium correlation between receptivity to engage in mobile information acquisition task and duration of \browse{s} (\spearmancorr{0.356}{0.052}), i.e., user receptivity increases when available duration increases. 
Participants shared in their interview that microlearning was better supported by \browse{s} that are longer than 3 minutes (\range{3}{6}, \meansd{4.3}{1.18}): \quoteby{P5}{I think I want to learn [3-6 words] when the duration is about 5 minutes [on the metro]}.  
As expected, receptivity towards microlearning was lower during \browse{s} with shorter durations. This depends on whether the value of switching to microlearning outweighs the switching cost, i.e., the time to switch from one task to another, including both physical and mental preparation time \cite{apa_multitasking_2019, katidioti_what_2014}. As described by some participants, \quoteby{P6}{[standing on the metro] I will get down in  2 minutes, it's not worthy to open the app}, \quoteby{P8}{[waiting for the bus] I need to get on the bus in a short time, maybe after I start, I can learn only a few words before I have to pay attention to the bus. So why not start learning after sitting down on the bus?}. 

In addition, the purpose and method of microlearning also mattered; participants highlighted that their preferred learning time depends on their personal learning technique, such as word repetition, making sentences or puns, and associating with objects/sounds. 

While the examples provided above are mainly based on microlearning, they can be extended to other information acquisition tasks. In general, \browse{} moments are best utilized for presenting information that is unrelated to their current activities. In addition, longer \browse{} moments are considered better for information acquisition than shorter ones. 

\factor{\Observe{}} 
\Observe{} is generally unsuitable for microlearning (\willing{2.4}{1.10}{237}) and other information acquisition tasks. However, receptivity levels differed based on two different types of engagement with the ongoing tasks; \i{essential engagement} and \i{inessential engagement}. As participant P3 summarized during an \i{essential engagement}, \quote{[chatting on social media] I don't want to learn at all. I want to chat with my friends and see their updates.} Similarly, frequent \observe{s}, like flipping through social media, texting friends, and intentionally looking at the road in crowded places, led to low receptivity towards other information acquisition tasks.  However, during \observe{s} related to \i{inessential engagements} which were carried out to \quote{kill} time (e.g., playing games, surfing the internet), or for tasks with low priority, participants showed greater receptivity to other information acquisition tasks: \quoteby{P1}{Oh, maybe it [observed \observe{}] is a good time to learn. I am just reading some gossip. You know when you are reading you won't think about language learning... I prefer doing more valuable things like vocabulary learning to refresh my mind.} The observed ratio between \i{essential} and \i{inessential engagement} was 224:13 = 95\%, indicating that users were mostly unwilling to engage in additional information acquisition tasks during \observe{} instances, as it disrupts their engagement with ongoing tasks. 

\factor{\Glance{}} 
As expected, \glance{s} due to their short duration and specific use are unsuitable for additional information acquisition tasks (\willing{1.2}{0.38}{58}). However, we noticed that \glance{s} interact with other visual behaviors to determine information acquisition suitability. During the commute, \glance{s} were needed in many navigational scenarios, and each \glance{} required the user to navigate their attention away from their current task. When the frequency of \glance{s} increased, users frequently experienced interruptions of their current task, which naturally leads to less efficient information acquisition. 
For example, when waiting for the bus to come, P2 was less willing to engage in an information acquisition task due to a lack of proper estimation for their bus arrival time, \quote{I think the bus is coming soon, but I am not sure. I have to check from time to time. [it] is too distracting}. According to our observations, the frequency of \glance{s} increased when there is unfamiliarity with place/road, uncertainty with the surroundings, vehicle motion (e,g., sudden jerks), and \envinterrupts{} (i.e., dynamic interruptions from the environment). For example, P3 stated, \quote{if I see people are coming, I need to look at them more often to avoid bumping. Then I don’t want to learn.}

\factor{\Envinterrupts{}}
Overall, there were two types of dynamic signals/cues from the surrounding environment that led to \glance{s} and potentially interrupted information acquisition: \signalforperception{s} and \signalforaction{s}. \Signalforperception{s} lead to the perceiving or cognitive processing of the cue without resulting in any associated action, while \signalforaction{s} lead to user actions (see Table~\ref{table:env_signals}). For instance, the scene outside the bus is a \signalforperception{}, while a change in the traffic light is a \signalforaction{} when the participant is waiting to cross the road, as it calls for action.

The frequency of perceived \signalforaction{s} and \signalforperception{s} was highly dependent on the time of the day, route, posture, and medium of commute. 
For example, the \signalforaction{ (flocking in/out of a crowd)} frequency while commuting by bus or metro during weekdays happened more frequently in the morning ($\approx$ 1 signal per 5-15 min) than at noon ($\approx$ 1 signal per 30-60 min).

\begin{table*}[h]
\caption{\Signal{s} from the surroundings}
\label{table:env_signals}
\begin{tabular}{ll}
\toprule
\textbf{\Signalforperception{s}} & \textbf{\Signalforaction{s}} \\ \midrule
1 or 2 passengers getting on/off the vehicle            & The flocking in of a crowd of passengers      \\
The movement of a few people on the road               & The movement of a crowd of people on the road \\
The fluctuation of the bus                             & The color change of the traffic light         \\
The slight changes of people's postures who are around & The arrival of the metro or the bus           \\ 
Announcements not related to commuting routes             & Warning sounds on the metro-station \\\bottomrule
\end{tabular}
\end{table*}

\subsubsection{Factors affecting the utilization of visual behaviors for information acquisition tasks}
\label{sec:study1_factors_consideration}
In summary, we identified four \i{user-related} factors to consider when adapting opportunistic visual behaviors in dynamic environments for information acquisition, such as microlearning.

\begin{enumerate}
    \item \textit{Balancing between situational awareness and information acquisition}.
    When \signal{} frequency or intensity increases, users would be less likely to engage in additional information acquisition tasks. Instead, they would prefer to focus on the surroundings only. \quote{I have to be more concentrated on the surroundings now. If I continue looking at my phone, I am afraid of bumping into others}, said P3 during a crowd influx. Since users have to act on \signalforaction{s}, ongoing information tasks should be paused when such signals are detected to allow users to focus on \envinterrupts{}. Similarly,  when any \signal{} is present, delaying the sending of new information tasks until users can focus on information tasks reduces the need to switch between \signal{s} and tasks, hence also reducing perceived information overload.

    \item \textit{Managing switching costs between navigational tasks and information acquisition}. 
    Participants expressed that they need a minimum available time before considering task switching. The amount of time required depends on the type of information acquisition task. In the case of microlearning vocabularies, participants expressed that they need at least 3 to 6 minutes to learn a few words. Hence, the perceived switching cost impedes the utilization of short \browse{s} for information acquisition, given that actual acquisition can occur in a lesser duration (e.g., less than 20s is required to associate 1-word pair \cite{dingler_language_2017}). The higher perceived switching costs also depend on other factors, such as the time taken to start the mobile app after taking it out of the pocket and unlocking it. Thus, in reducing switching costs, we can enable the utilization of shorter durations for productive information tasks.

    \item \textit{Balancing between the ongoing cognitive processes and information acquisition}. 
    Even though \browse{s} are the most opportunistic visual behavior for information acquisition, when users are engaged in other cognitive processes (i.e., \activebrowse{s}), they become unwilling to engage in microlearning, \quoteby{P7}{[sitting on the bus, during observed \browse{}] I had a class. I need to relax, I don't want to learn now.} The presentation of information tasks during \activebrowse{s} should be avoided, so as to minimize any potential annoyance to the user.
    
    \item \textit{Difficulties in self-identifying opportunistic moments for information acquisition}. 
    Since their own behavior patterns were not observable to participants, they did not engage with \quote{valuable} activities even if they were more receptive. This indicates a need for developing methods of identifying and providing feedback about potential opportunities (e.g., \passivebrowse{s}).  
\end{enumerate}

\subsubsection{Device-task interplay}
\label{sec:mobilephone-limitations}
From the inquiries, we identified two mobile phone limitations for microlearning.
\begin{enumerate}
    \item The reduction of peripheral vision field when visually focused on phone screens. This visual field constriction made it difficult for participants (\participantproportion{7}{8}) to maintain situational awareness while learning with their phones during commuting. As P3 mentioned, \quote{Sometimes I am focused on finishing a [mobile learning] session, and I can not see when somebody needs me to move out of their way.}
    
    \item Ergonomics. \participantcount{4} complained about the heads-down posture and associated fatigue with mobile phones, \quoteby{P4}{Focusing on the phone is really tiring. I need to nod down to look at the screen, and it hurts my neck.}  
\end{enumerate}
Additionally, three participants did not prefer using mobile phones for learning vocabulary while commuting due to their propensity to motion sickness, \quoteby{P6}{I can not focus [a] long time on [the] phone especially on [the] bus as I feel carsick.}

Existing literature supports the device limitations that we identified.
Visual field constriction \cite{maples2008effects} is a key factor in the `smart-phone zombies' phenomena \cite{appel_smartphone_2019}, which describes mobile phone users who are obsessively engaged with their phones and compromise on situational awareness \cite{lin2017impact, basch_pedestrian_2015}. Moreover, detrimental long-term effects of the head-down posture include health problems such as the `Text Neck' \cite{textneck, gustafsson2017texting}, a form of chronic musculoskeletal disorder.

\subsection{Discussion}
\label{sec:study1_discussion}

Different visual behavior patterns cater to the specific needs of different tasks. Patterns transition when users switch tasks, self-interrupt (e.g., choosing to relax after interactions), or react to \envinterrupts{} (e.g., \signal{s}). 

User receptivity to information acquisition depends on the behavior pattern and perceived attention needs. The Resource Completion Framework (RCF) \cite{oulasvirta2005interaction} describes the cognitive resource allocation during mobile HCI tasks, which we can use to explain user receptivity levels during each visual behavior pattern.
According to the RCF, when an information task dominates one's attention, the working memory processes retained information, increasing the cognitive load. The cognitive load affects the perceived demand of tasks, the effort required to start or continue a task, and users' receptivity towards the information task. We found that \Browse{} requires the least visual attention; thus, users have a greater capacity to divert their attention to other information acquisition tasks. In contrast, \observe{s} were poorly received since it is attention-demanding. Therefore, we can predict opportune moments for information acquisition tasks and forecast user receptivity by observing their visual behavior patterns.

It is essential to consider the threshold or lower bound time limit for effective information acquisition. The effective association of a word pair takes at least several seconds\cite{cai_wait_learning_2015}, suggesting that visual behaviors shorter than $\approx$ 8s (minimum time required for users to associate a word pair \cite{dingler_language_2017}) are unsuitable for microlearning. Some \glance{s} and very short \browse{s} are too brief for most information acquisition tasks. 
Visual behaviors of shorter durations (e.g., \glance{s}) impact users' receptivity, as the frequency of such behaviors also affects the frequency of transitions. 
This is because \glance{s} involve checking and remembering problem-states, i.e., working information relevant to the ongoing task \cite{borst_2015_interrupts}. Hence, when the prediction of the problem-state is inaccurate, more frequent \glance{s} are required, requiring greater effort from the user to store and restore multiple problem-states. This cognitive demand may lower user receptivity.

According to our study, \browse{s} of longer durations are common and useful. In the case of microlearning, as much as 65.4\% of observed \browse{s} are longer than 8s. These account for 92.3\% of the total \browse{ing} time. Even with a higher duration threshold of 33s (median duration), there will still be  50.9\% of the \browse{ing} time (83.7\% of the total duration). 
This provides considerable opportunity for information acquisition tasks since \browse{s} account for more than one-fifth of commuting time (21.8\% of total commute time). Furthermore, more than 60.8\% of \observe{s} durations are carried out for mobile-HCI tasks (each duration lasting longer than 33s), where some of them are carried out for nonproductive tasks. While our observations indicate many opportunities for information acquisition on the go, it is vital to avoid overusing these opportunities to the point of information overload and mental fatigue. 

\section{Study 2: Opportunities with visual behaviors for microlearning with mobile phones and OHMDs}

\Studyone{} established the visual behaviors that commuters engage in and the desirability of utilizing them for information acquisition tasks. However, since we did not test the actual usage in any mobile platform, the results were not validated ecologically. 

We proceed to investigate visual behaviors and information acquisition opportunities on different platforms, particularly by comparing the de-facto mobile phone platform with the emergent OHMD platform. Both platforms are designed for mobile usage but have distinct characteristics that can provide different information acquisition opportunities. 
\vspace{1mm}

\noindent{}Our research questions for \Studytwo{} includes:
\vspace{1mm}

\noindent{}\b{RQ1:} How are visual behaviors for microlearning utilized differently on the mobile phone and OHMD platforms? How do the platforms affect user receptivity to additional information acquisition tasks (i.e., microlearning)?
\vspace{1mm}

\b{RQ1.1}: What are the limitations of each platform in utilizing visual behaviors for information acquisition (e.g., microlearning)?
\vspace{1mm}

\noindent{}To answer these research questions, we probed \cite{hutchinson_technology_2003} microlearning on both mobile phones and OHMDs separately, focusing on \browse{s} that users were most receptive to. Probing allowed us to focus on identifying the relationship between device platforms and user-related factors identified in \studyone{} (sec~\ref{sec:study1_factors_consideration}) without being subject to the technical constraints of an actual implementation.
Therefore, we used the \i{push} strategy \cite{isaacs_mobile_2009} to \b{remind} users of potential microlearning opportunities and verify whether the platforms (mobile phone and OHMD) enable users to utilize them and identify the associated tradeoffs. 

\subsection{Probe study design}
\label{sec:apparatus}

To make the information acquisition task more realistic, we created a vocabulary list for microlearning, similar to what is presented in current vocabulary learning apps such as Duolingo\footnote{\url{https://www.duolingo.com/}}, which contains both visual (spelling) and auditory (pronunciation) cues. To ensure that no participants have prior experience with the selected vocabulary list, we used Vimmi corpus \cite{macedonia_vimmi_2010}, an artificial corpus, as the second language.
We created 90 Vimmi-English word pairs (e.g., ``toze'' in Vimmi refers to ``flower'' in English) as part of the microlearning material.

\begin{figure}[h]
  \centering
  \includegraphics[width=\linewidth]{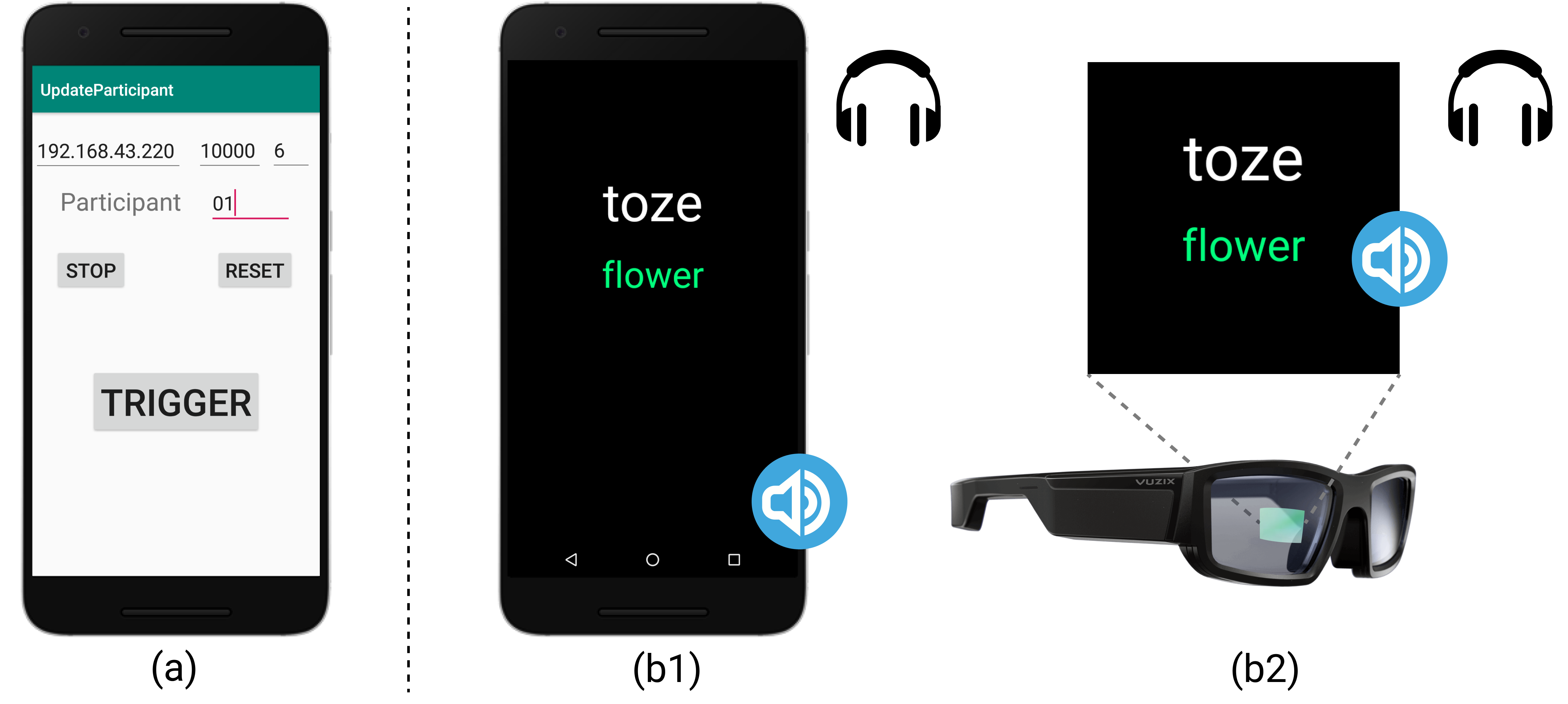}
  \caption{Apparatus used in the probe-based study design: (a) Mobile application interface to trigger the microlearning session. The participant's device (mobile phone or OHMD) IP address, microlearning session parameters (e.g., 10000ms time gap between words, 6 words per session), and participant id were configured during the training. The microlearning session was triggered by clicking the `TRIGGER' button. (b1) Mobile microlearning application interface (b2) OHMD (Vuzix Blade) microlearning application interface. The trigger displayed the Vimmi-English word pairs and played the audio pronunciation in the device. NOTE: In OHMDs, the black color background represents the transparent background.}
  \Description{Apparatus: (a) Mobile triggering application interface (b1) Mobile microlearning application interface (b2) Smart glasses microlearning application interface}
  \label{fig:study2_appratus}	  
\end{figure}

We designed two Android mobile apps, one for participant microlearning (Fig~\ref{fig:study2_appratus} (b1), (b2)) and the other for the experimenter to trigger microlearning (Fig~\ref{fig:study2_appratus} (a)) in the participant app. Whenever the experimenter triggered a microlearning session, Vimmi-English word pairs were automatically displayed and audio pronunciation sounded on the participant's device (see Fig~\ref{fig:study2_appratus} (b1), (b2)). 

Each microlearning session included 6-word pairs, a design choice based on previous study results \cite{dingler_language_2017}. To determine the display duration of each word pair, we conducted a pilot on three volunteers, where they had to learn vocabulary on both platforms on the move. All participants preferred the duration of 10 seconds. 

While learning involves different stages (i.e., acquisition, retention, and transference of knowledge \cite{ausubel_acquisition_2000}), we focused on the \b{acquisition stage} as it is the first step and fundamental to the other stages of retention and transfer. Therefore, we did not adopt any vocabulary learning techniques (e.g., spaced repetition).

In our informal pilot study, participants detected words shown on their OHMD without the need for any additional notification though this was not the case for mobile phones. Therefore, mobile phone words were triggered with vibration and a 1s audio beep to notify participants (who might have had their attention off-screen) that a word had been triggered. For both platforms, words were shown on screen for a fixed duration before disappearing. Thus, users who chose not to pay attention missed out on the displayed content.

\subsection{Method}

\subsubsection{Apparatus}
For the \phoneprobe{}, participants installed the microlearning mobile app (Fig~\ref{fig:study2_appratus} (b1)) on their own phones. For the \glassprobe{}, participants wore a pair of Vuzix Blade\footnote{\url{https://www.vuzix.com/products}} smart glasses (480x480 px display, centered on the right glass as shown in Fig~\ref{fig:study2_appratus} (b2)), installed with a customized version of the mobile app (see sec~\ref{sec:apparatus}). In both probes, participants wore earphones to listen to the audio. The experimenter used a Google Pixel 4 phone to trigger the microlearning sessions.

\subsubsection{Participants}
A total of 20 volunteers were recruited for the two probes (12 females, age \meansd{23.6}{3.1}, \range{19}{30}). 16 of them were students, 3 were IT professionals, and 1 was a business professional. All participants had received formal education in English and had experience using mobile learning apps during commuting. However, none of them had prior experience using OHMDs. In addition, all participants were regular commuters, spending an average of 131 minutes (\range{80}{300}) per day commuting.

We recruited 16 participants with a between-subject design to obtain diverse feedback and minimize possible interference between the two conditions. Eight (P1-P8) participated in the \phoneprobe{}, and eight (G1-G8) participated in the \glassprobe{}. We balanced participants between the two probes based on their commuting time and medium. While between-subject design eliminates the possible interference between conditions, it is less sensitive to detect subtle differences between conditions due to subjective differences. To compensate for this, we also tested 4 more participants (\b{GP1-GP4}, 1 IT-professional, 3-students) with a within-subject design in which all 4 participants underwent both probes. 

\subsection{Procedure}
\begin{figure*}[h]
  \centering
  \includegraphics[width=\textwidth]{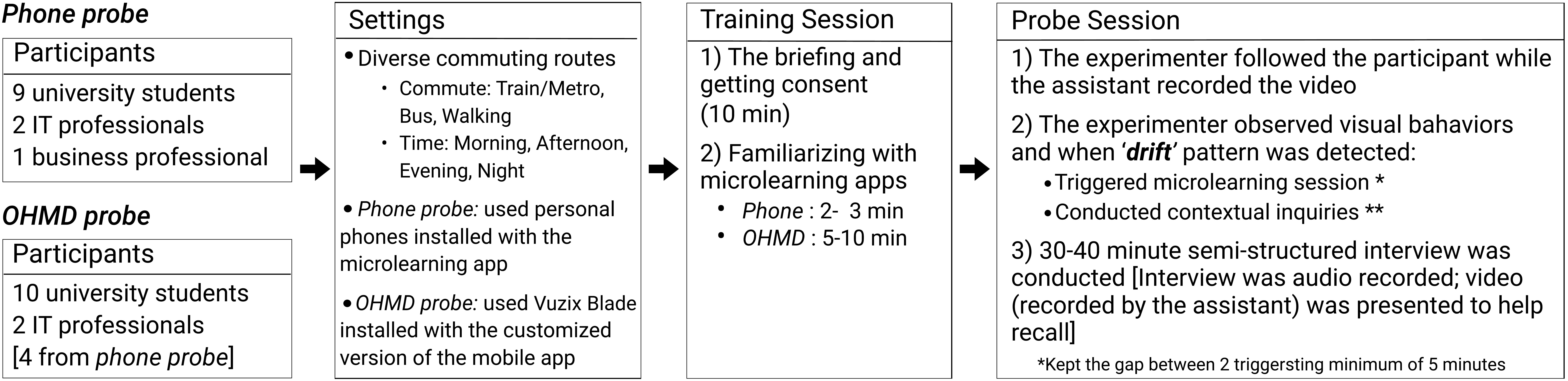}
  \caption{Study 2 procedure. See Appendix~\ref{appendix:study2_procedure} for further details.}
  \Description{Study 2 procedure. See Appendix~\ref{appendix:study2_procedure} for further details.}	
  \label{fig:study2_procedure}	  
\end{figure*}

Fig~\ref{fig:study2_procedure} illustrates the process of the probe study (for details, refer to Appendix~ \ref{appendix:study2_procedure}). 
We did not test the number of words participants remembered (i.e., information \i{retention}) as the objective was to identify how the two platforms support information \b{acquisition} of new concepts and \b{receptivity} of platforms.

\subsubsection{Measures}
\label{sec:study2:measures}
Similar to \studyone{}, the primary measure was participants' feedback. A 5-point Likert scale (1 = Very Low, 5 = Very High) was used to gather participants' receptivity and concentration on microlearning. The contextual inquiry topics and interview topics can be found in Appendix~\ref{appendix:study2_contextul_inquiry} and Appendix~\ref{appendix:study2_interview}.

\subsection{Data Analysis}
We conducted a qualitative analysis with 247 sessions of contextual inquiry notes, observation notes, and 24 interview transcripts. Two researchers (co-authors, one experimenter) used the \textit{QDA Miner} software package and analyzed notes using the open coding technique \cite{corbin_grounded_1990, corbin2014basics}. An initial set of common codes were generated by analyzing 4 participants' data (2 per platform). This initial set of codes were further refined after watching video recordings and further discussion. The refined coding scheme was then iteratively tested and revised to reach 94\% agreement among the coders before coding the rest of the participants.  


\section{Findings: Study 2}
\label{sec:probe-result}
All participants engaged with microlearning during the observed 247 \browse{s}, except for 6 technical fault cases where the microlearning app failed to start upon triggering. We compared the overall probe duration and session count statistics to verify if participants had similar experiences with the two platforms. On average, \phoneprobe{} participants spent 70 minutes (SD = 25), while \glassprobe{} participants spent 73 minutes (SD = 32) on commute during the study, excluding briefing, training, and post-interview time. On average, \phoneprobe{} participants received 9.2 sessions/hour (SD = 1.6) while \glassprobe{} participants received 9.4 sessions/hour (SD = 1.5). Additionally, we observed that between-subjects participants had similar feedback on their experiences as within-subject participants. 

We organized the results based on the research questions and themes that emerged from our data analysis.

\subsection{RQ1: How are visual behaviors for information acquisition utilized differently on the mobile phone and OHMD platforms?}
Overall, we found that the receptivity to information acquisition (microlearning) primarily depended on \envinterrupts{} (i.e., \signal{s}, sec~\ref{subsec:behavior_and_receptivity}). However, the platform also had a strong influence on how these opportunities were utilized. 

\subsubsection{\Envinterrupts{}, microlearning, and situational awareness.}
\label{sec:interrupt and signal}
\begin{figure}[h]
\centering

\begin{subfigure}{.45\textwidth}
  \centering
  \includegraphics[width=\linewidth]{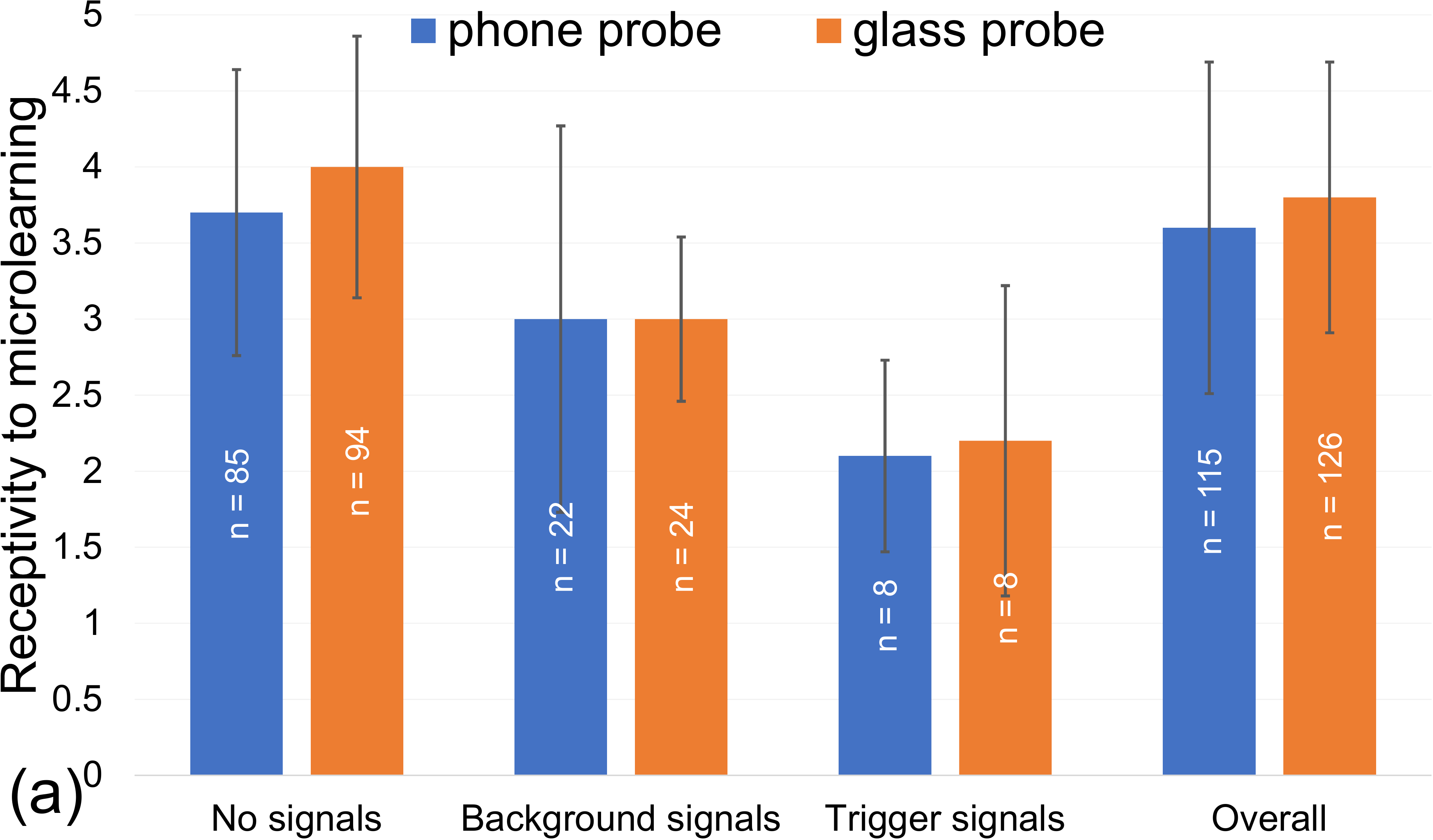}
\end{subfigure}
\begin{subfigure}{.45\textwidth}
  \centering
  \includegraphics[width=\linewidth]{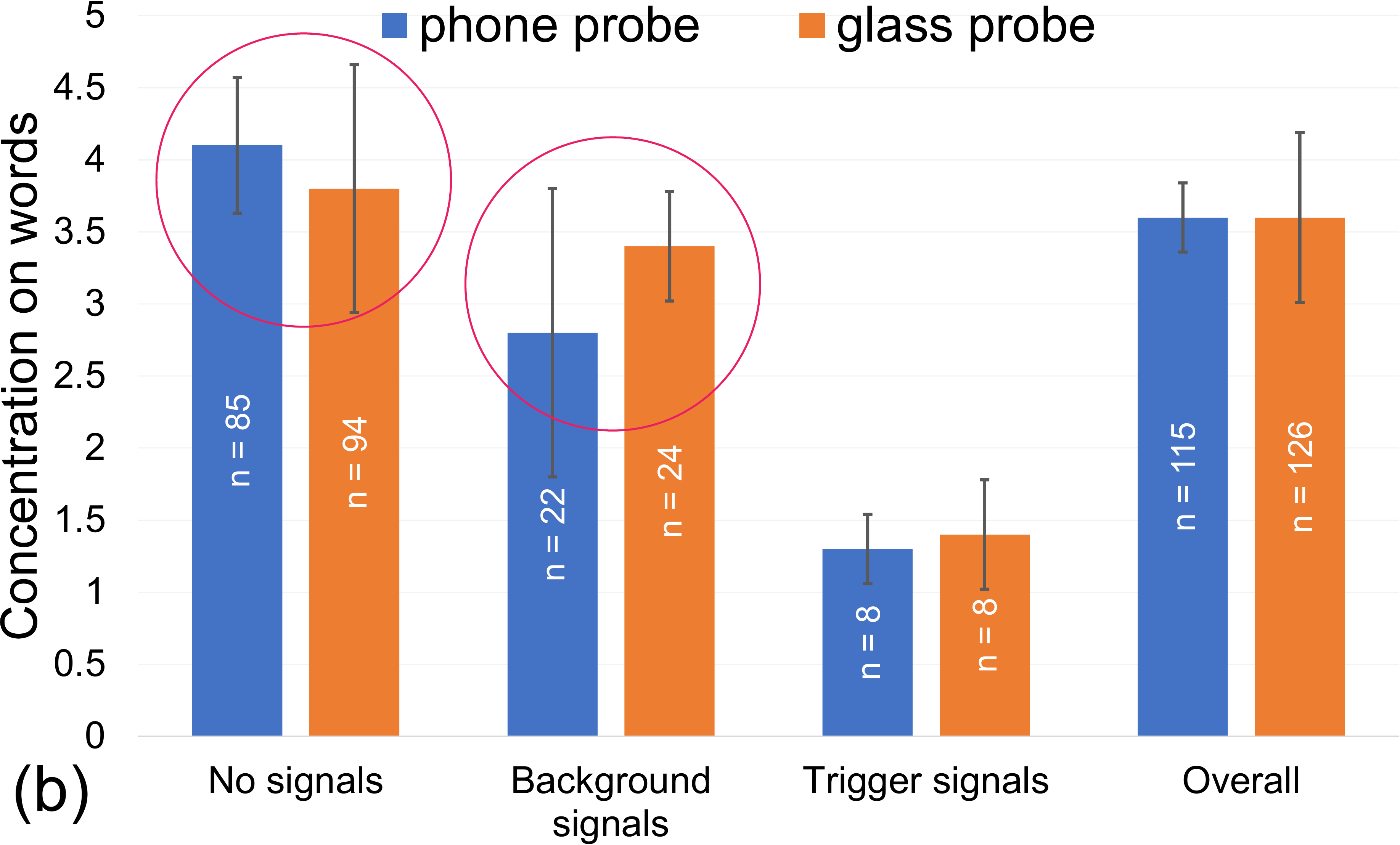}
\end{subfigure}

\caption{\Signal{s'} effect on $(a)$ receptivity  and $(b)$ concentration  on information acquisition (microlearning). The Y-axes of both graphs represent 5-point Likert scales, and error bars represent the standard deviation. NOTE: \Signal{s} occurred after the information acquisition task was triggered during \browse{s}.}
\Description{Two bar charts representing the perceived behavior of participants. Chart (a) shows the receptivity to microlearning during phone probe and glass probe under 4 conditions; no signals, with background signals, with trigger signals, and overall. Chart (b) show the concentration on words during phone probe and glass probe under 4 conditions; no signals, with background signals, with trigger signals, and overall.}
\label{fig:platforms_signals}
\end{figure}

In general, participants' receptivity to engage in microlearning, as well as their concentration levels reduced as the \signal{} (sec~\ref{subsec:behavior_and_receptivity}) intensity increased (Fig~\ref{fig:platforms_signals}, \added{compare \i{no signals} or \signalforperception{s} with \signalforaction{s}}). However, the two platforms differed in this aspect. As shown in Fig~\ref{fig:platforms_signals}a, the \i{overall} receptivity did not vary much across platforms. This is because users' receptivity depended on their information needs and motivations.

\begin{figure}[h]
  \centering
  \includegraphics[width=\linewidth]{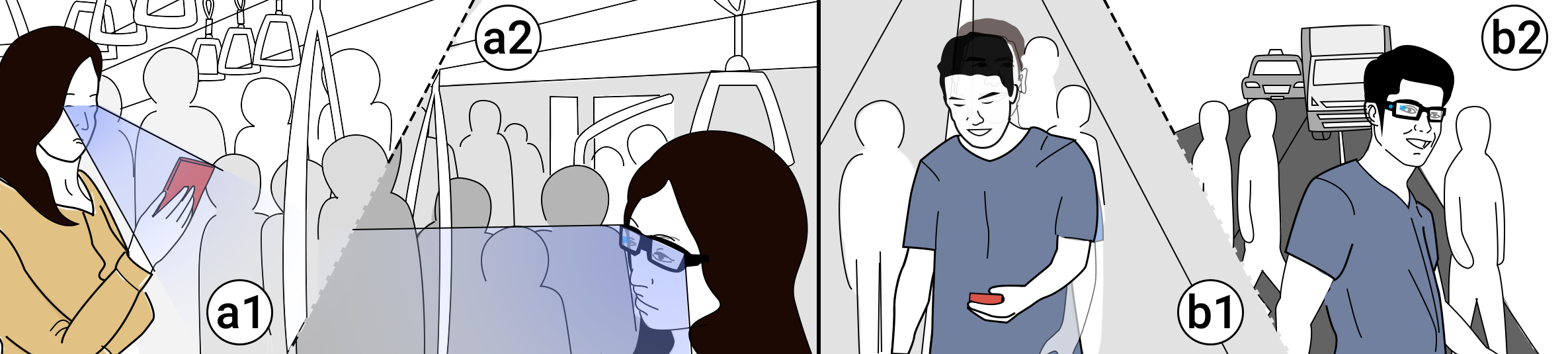}	  
  \caption{Situational awareness on mobile phones and OHMDs: (a1) The mobile phone user needs to lift the phone to pay peripheral attention to the crowd; (a2) The OHMD user looks ahead and focuses on the words while paying peripheral attention to the crowd; (b1) While walking, the person needs to look up to check the surroundings and look down to focus on the phone; (b2) While walking, the OHMD user can check the road while paying minor attention to words}
  \Description{Situational awareness on mobile phone and OHMDs under traveling on bus and walking on road.}	
  \label{fig:situational_awareness}	  
\end{figure}

As shown in Fig~\ref{fig:platforms_signals}b, when there were few \signal{s}, mobile phones enabled better concentration on information acquisition due to their static (opaque) backgrounds and high-resolution screens. On the other hand, when there were more \signal{s}, information acquisition on mobile phones was more frequently interrupted, significantly affecting participants' concentration. In contrast, OHMDs supported better concentration as users tended to be more situationally aware of \signal{s} in most commuting scenarios (refer to Fig~\ref{fig:situational_awareness} (a2), (b2)). 

All OHMD users checked their surroundings through the see-through displays and used their peripheral vision to monitor \signal{s} and decide on the type of signal: \signalforperception{} or \signalforaction{}. Their microlearning was only interrupted during \signalforaction{s}: \quoteby{G1}{[waiting for the bus, \signalforperception{}] I don't have to change my gaze to check the bus deliberately. I know clearly whether the bus is coming even when my eyes [are] still glued to the words.}, and \quoteby{G3}{[walking in the metro station, suddenly a child runs in front, i.e., \signalforaction{}] I managed not to bump into the kid even though I was looking at words, but I lost my focus}. The see-through nature of OHMDs helped users to avoid situational hazards, allowing them to pay attention to information acquisition and dynamic \signal{s}.

In contrast to OHMDs, mobile phones were not designed for users to effectively acquire information on screen while monitoring the environment. 

There were also disadvantages to using peripheral vision for OHMD content and focal vision for the surroundings. When \participantcount{3} had to pay more attention to the \signal{s}, words that appeared in the peripheral vision were distracting: \quoteby{G2}{[waiting for the bus] Actually, I was quite okay with learning when I was checking the bus. But now [walking towards the bus], I find the words really distracting. I can't help looking at the words from time to time, but I become quite worried about bumping into others when I do it.} This indicates that information acquisition tasks on OHMD are best paused when \signal{s} require the users' focused attention.

\subsubsection{Learning strategies and behavior changes.}
\label{sec:interrupt_and_learn}
There were differences between the two platforms regarding how users engaged in information acquisition and how it affected their visual behavior.

OHMD users could adopt a \covertlearning{} strategy (\participantcount{8}) when they had to pay more attention to the surrounding information (e.g., \signal{s}), which enabled a more continuous learning experience compared to mobile phones. A \covertlearning{} strategy refers to the use of peripheral vision to check on learning content while focal vision is occupied by \signal{s}. For example, according to G2, \quote{When I was checking the number of the bus, I can still see the words in the range of vision.}. G1, on the other hand, described it as a ``half-look'': \quote{even when I am looking elsewhere, I can `half-look' at the words at the same time.} Moreover, this \covertlearning{} strategy can minimize interruption to microlearning with frequent \glance{s} in between \browse{s}, and is suitable for use during commute when attention is often fragmented. Given that \envinterrupts{} are unavoidable on the move, \covertlearning{} can reduce the cost of context switching between information acquisition and attending \signal{s} and improve the efficiency of information acquisition. We suggest future investigations for validating this hypothesis.

Mobile phone users could hardly leverage \covertlearning{} since keeping their phone screen within the same range of vision as when their gaze is directed towards their immediate environment was challenging. Instead, all participants used these moments away from the screen to digest and memorize words: \quoteby{GP3}{When I look down at the phone, I am intaking the word, just to leave a rough impression. When I look up and check the road, I can digest the word and try to memorize it. After checking the road, I refocus on the word for [a] deeper impression.} For this case, we recommend further studies to identify the most effective strategy for information \i{retention} between two platforms on the move. 

Information acquisition tasks clearly influenced visual behaviors. In the case of microlearning, participants switched from \browse{} to \glance{} (check the word) or \observe{} (recognizing the word) and then back to \browse{} (trying to memorize words). In the \phoneprobe{}, all participants used to look \i{down (word) and up (environment)}; while in the \glassprobe{}, all participants used to \i{focus (word) and defocus (word)}, to check and remember words: \quoteby{G8}{I focus on the words [be]cause I saw it, but then when it [word] changes then I'll refocus on the change.} While mobile phone users explicitly used both head movements and gaze to focus and defocus on words (e.g., focus on environment), OHMD users only used gaze to do so.

\subsubsection{Switching cost.}
\label{subsec:study2:switch_cost}
Overall, OHMDs can lower switching costs as less physical effort and gaze time are required of users. 
OHMDs require shorter transition times \quote{fraction of second} ($\approx$ 150 ms according to \cite{rayner_eye_2009}) than phones (we observed a 2-6 seconds transition time). The ease in transition considerably lowers the psychological barrier towards information acquisition tasks, as G4 commented: \quote{the smart glasses quickly bring me into a learning state without any delay. But usually, when I receive notifications from the phone, I need more time to switch into a good learning condition. Maybe I will wait for a while before picking it up. Or maybe I don't bother.} 

Due to low switching costs, OHMDs could utilize shorter chunks of time for information acquisition and thus offer more opportunities for information acquisition tasks than mobile phones. For example, in certain places such as the escalator where participants had 40-60 second \browse{s}, most participants (\participantproportion{7}{9}) showed low receptivity to the \phoneprobe{}. P4 highlighted this, \quote{the time is very short, so after I picked up the phone and started learning, maybe I have to put it down and prepare for getting off before I can remember a single word. I'd rather do nothing but wait.} On the contrary, participants were more receptive during short \browse{s} in the \glassprobe{} (\participantproportion{7}{8}): \quoteby{GP2}{I think even 10 seconds is possible to be utilized for smart glasses [to microlearn]. I just need to quickly glance at the words and then quickly defocus to prepare for the next task.}

However, for users who are not ready to accept information, the low switching costs of OHMDs could cause unwanted interruptions to ongoing mental processes. For example, two participants complained that their thought processes were interrupted by OHMD content that appeared, \quoteby{GP4}{I was thinking something important just now. But the words suddenly appeared, and I forgot what I was thinking}. On the other hand, one participant from the \phoneprobe{} missed two sessions as he did not notice the vibration or the audio beep while he was pondering. Therefore, OHMDs are more likely to result in accidental triggers that potentially annoy users. 

\subsubsection{Ergonomics, postures, and motion effects.}

OHMDs could mitigate the ergonomics issues identified in \studyone{} (sec~\ref{sec:mobilephone-limitations}) by supporting the natural heads-up posture. It could also reduce the perceived effects of vehicle motion, reduce disturbances to information acquisition, and improve receptivity.
Three participants in the \phoneprobe{} expressed that looking up and down was \quote{tiring} or \quote{annoying} when they were microlearning. In contrast, five participants in the \glassprobe{} expressed that OHMDs were \quote{much more comfortable} than mobile phones (based on their past experience) and thus, increased their willingness to learn. 

Moreover, we observed the influence of posture in the \phoneprobe{}: \participantcount{5} changed their posture from standing to sitting with no \signal{s}, and as a result became more receptive (\willing{3.5}{1.02}{42} \b{to} \willing{4.4}{0.41}{32}). P1 mentioned in the \phoneprobe{}, \quote{when I am sitting down, I feel more comfortable. I don't move my head so often even [...] my willingness is higher [compared to standing]}. Yet, three participants mentioned that they preferred standing over sitting due to the potential \signal{s}. For example, P5 mentioned that \quote{I wouldn't have to keep looking up to check for people who need [my seat]}. In contrast, OHMD users were more ``relaxed'' as the OHMD supported personal preferences for posture as they engaged with microlearning. This made them more receptive to learning on OHMDs than on mobile phones during those particular scenarios. 

The see-through and view-stabilized nature of OHMD content reduced the adverse effects of vehicle motion (e.g., jerking) and sudden speed changes (\participantproportion{5}{7}). Mobile phone participants required more effort to keep their eyes on the screen in moving vehicles, which sometimes led to motion sickness. GP3 highlighted that \quote{[with the mobile phone, sitting on the bus] If it is bumping up and down, I need to look at the screen for a longer time to intake the information. And I will look up less frequently because it makes me sick.}, \quote{[with OHMDs] It's easy to focus on words [...] even with shaking}.
As suggested, OHMDs facilitate better concentration in more dynamic contexts (see Fig~\ref{fig:platforms_signals}b) and lower the barrier to information acquisition task engagement. In the next section, we expand on the limitations of each platform's use.

\subsection{RQ1.1: What are the limitations of each platform in utilizing visual behaviors for information acquisition?}

\subsubsection{Limitations of mobile phones}
All limitations identified in sec~\ref{sec:mobilephone-limitations} were also present in the \phoneprobe{} when participants physically experienced experimenter-triggered microlearning sessions during \browse{s}.

\subsubsection{Limitations of OHMDs}
\label{sec:smart_glasses-limitations}

\factor{Social limitations} 
Social context plays an essential role in participants' ability to acquire information. For instance, participants disliked having digital content appear on their OHMDs when they were also looking at the faces of others. In our study, all participants except one reported that they would first check that they were not interacting (directly/indirectly) with others before beginning to read their OHMD content. 

Focusing on OHMD content could lower participants' awareness of their social environment. Two participants encountered the issue of unintentionally staring at others when they were, in fact, focusing on displayed content: \quoteby{G6}{I suddenly realized that I was looking at someone's face when I looked away from the words, so I quickly changed the direction I was staring at. I am afraid that when I focus on the words, I will again unconsciously look at someone, so I became more careful.} In crowded environments, some would look at the ground or the ceiling to avoid such unintentional gazes at others; however, this \quote{unnatural} posture made learning \quote{uneasy}, an issue which OHMDs usually helps overcome. According to Akechi et al. \cite{akechi_attention_2013}, some cultural contexts consider gazes to be inappropriate. In cultures where this poses a sensitive social issue, OHMD users may be more inclined to ensure that they maintain a base level of situational awareness to avoid landing unintended gazes upon others. 

\factor{Technical Limitations}
We identified three areas of technical constraints for OHMDs that limit information acquisition on the move. The first constraint relates to the high external brightness and contrast, which is common in most optical see-through displays \cite{azuma_survey_1997, lucero_notifeye_2014, kerr_wearable_2011}. Eight participants mentioned that when the environment was bright, or the surrounding colors matched the text colors on display, they had to look for \quote{contrasting} surfaces to see OHMD words clearly. Three participants complained that reflections in the outdoor environments distracted them. The second limitation arises from the hardware properties of OHMDs. Four participants mentioned that OHMDs were heavy, inconvenient for use, and difficult to customize. The third limitation relates to the issue of eye strain \cite{han_comparison_2017, herzog_effects_2019}. Only one participant encountered this due to unfamiliarity with the OHMD prototype and its monocular nature: \quoteby{G2}{[I] feel the right eye is used more, even when the screen is centered}. 

Nevertheless, we anticipate that future advancements in OHMD technology will seek to resolve these technical limitations. For example, better projection technology will resolve issues of visual contrast and reflections \cite{glass_intel_2019}. We saw significant improvements in some of the recently released OHMDs, such as the Nreal\footnote{Nreal light: \url{https://www.nreal.ai/light}} smart glasses, which is comfortable to wear and delivers a better viewing experience.

\subsection{Summary of the findings}
Table~\ref{table:summary_findings} summarizes the key findings of this study.  

\begin{table*}
\caption{Summary of the key findings. \plus{} indicates positive qualities, \minus{} indicates negative qualities, and \neutral{} indicates neutral qualities.}
\label{table:summary_findings}
\begin{tabular}{p{0.49\textwidth} p{0.49\textwidth}}
\toprule
\b{Mobile phones} & \b{OHMDs / Smart glasses} \\ \midrule
\plus{} Allow users to better concentrate on the information acquisition task when no/fewer \signal{s} exist & \plus{} Allow users to better concentrate on the information acquisition task even when \signalforperception{s} are present \\ \midrule
\minus{} Limit the user's ability to be situationally aware and engage in the information acquisition task due to visual field constriction & \plus{} Allow users to have situational awareness and engage in the information acquisition task \\ \midrule
\minus{} Higher switching cost between the environment and task due to the greater physical effort required & \plus{} Lower switching cost due to direct gaze interaction \\ 
 & \minus{} Can interrupt cognitive processes when sudden visual stimuli (e.g., digital content) appear in front of eyes \\ \midrule
\neutral{} Require users to look up and down to memorize words & \neutral{} Require users to focus and defocus to memorize words \\
 & \plus{} Can use \covertlearning{} to continue information acquisition \\ \midrule
\minus{} Can cause neck fatigue due to the head-down posture during interaction & (+) Do not cause neck fatigue as the head-up posture is more relaxed \\ \midrule
\minus{} Can be challenging to concentrate on words with vehicle vibration and motion  & \plus{} Easy to concentrate on words with vehicle vibration and motion \\ \midrule
 & \minus{} Can be socially awkward when words appear in front of faces \\
 & \minus{} Can cause eye strain when users are not accustomed to OHMDs. More training is required to increase familiarity levels. \\
\bottomrule
\end{tabular}
\end{table*}
\section{Discussion and Design Implications}

We aimed to understand the on-the-move visual behaviors and utilize them to support mobile information needs on different mobile platforms. Using a shadowing study in the wild, we identified 3 visual behaviors patterns: \glance{}, \observe{}, and \browse{}. Subsequently, we probed on OHMDs and mobile phones in the wild to determine the influence of device platforms on microlearning opportunities created by \browse{} behavior. We found that the OHMD platform provides more opportunities for mobile information acquisition, while the mobile phone platform facilitates a limited yet more focused information acquisition experience. In the next section, we discuss methods for supporting information acquisition on the move, then, more specifically, the use of OHMDs for this purpose.

\subsection{Design for information acquisition on the move: Information acquisition vs. Navigation}
\subsubsection{Detection of `opportunistic' \browse{s}}
\label{subsec:opportunistic_browse}
We gained two key insights on opportunistic behaviors: (1) Out of the three visual behavior patterns, \browse{s} provide the most opportunistic moments for mobile information acquisition, (2) Not all \browse{s} present opportune moments. 
Since \browse{s} are mostly of sufficient duration for information acquisition (sec~\ref{sec:visual_behavior_patterns}: 83.6\% in duration above the median, or 22\% of commuting duration), we propose a way to detect and utilize opportunistic \browse{s} for information acquisition tasks.
The first step involves detecting \browse{s}. This can be achieved through mobile gaze tracking such as via object/location of eye focus \cite{zhang_everyday_2017}, head orientation \cite{wtiefelhagen_head_2002}, and alertness detection \cite{tag_continuous_2019}. After the \browse{} visual behavior is identified, the second step involves filtering out moments when users are engaged in nonvisual cognitive tasks (i.e., \activebrowse{}). We can achieve this through cognitive load monitoring with physiological sensing of pupil size, ECG, or EEG \cite{haapalainen_psycho_physiological_2010, iqbal_index_2005, brouwer_2012, berka_eeg_2007}.

\subsubsection{Review vs. New}
Since different opportunistic \browse{s} are of different durations, information acquisition tasks should be assigned based on the available duration so as to avoid overloading users with information. 
For example, tasks involving word-learning are not suitably achieved with short \browse{s} as they require time and concentration. All participants in \studytwo{} preferred reviewing materials they had previously seen instead of learning new ones during short \browse{s}. As expressed by P2: \quote{[before boarding the bus] The interval is too short for me to intake new information. I think I'd rather review old words now since I need less time for each word.}

However, it is difficult to predict the duration of \browse{s} due to the dynamic nature of \signal{s}. It is thus helpful to assign a shorter information acquisition task (e.g., review session) once a \browse{} has been identified, regardless of its predicted duration. As elucidated in sec~\ref{sec:interrupt_and_learn}, when users tried to memorize words, they switched from \browse{s} to a series of \glance{s} and \observe{s}. In this case, the system should initiate a longer information task (e.g., a learning session with new words) upon detection of user behavior from a previous information acquisition task. \added{If there are unexpected \signal{s} from the environment, users may switch away from word-memorizing behaviors, and such deviations can be used to stop/pause information tasks.}

\subsubsection{Automatic vs. Manual control}
\label{subsec:automatic_manual}
Automatic pushing has many advantages: participants do not have to retrieve information manually, and the pushed content reminded and encouraged them to take on the information acquisition opportunities available from their devices. Despite these benefits, some participants considered automatic pushing to be inappropriate and annoying, primarily when they were already engaged in something else (e.g., pondering, relaxation, sec~\ref{subsec:study2:switch_cost}). Whenever their environmental conditions were changing and uncertainty high, users frequently switched their attention between information acquisition and commute-related tasks, preferring to eliminate acquiring digital information. 
Thus, we suggest automatic pushing for kick-starting information acquisition (e.g., microlearning), then allowing users to manually postpone or cancel the session through peripheral interactions if preferred. We present these suggestions in a flowchart, Fig~\ref{fig:automatic_detection} (see Appendix~\ref{sec:appendix:automatic_detection} for the proposed solution for microlearning). 

\begin{figure}[h]
  \centering
  \includegraphics[width=\linewidth]{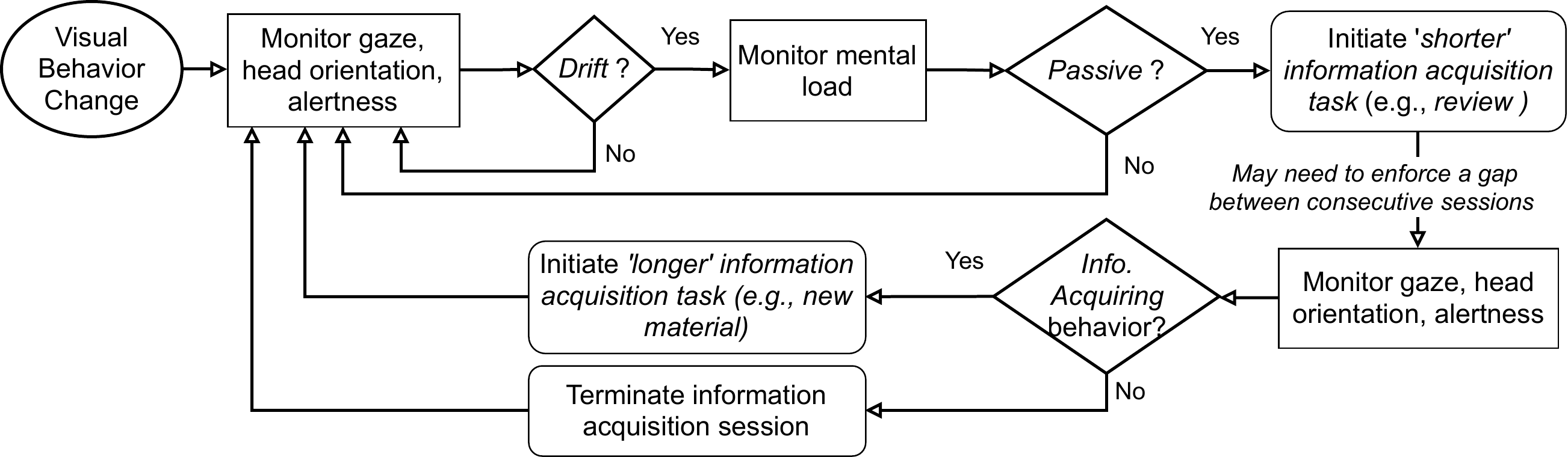}
  \caption{A flowchart for automatic detection and utilization of opportunistic \browse{s} for information acquisition. \added{See Appendix~\ref{sec:appendix:automatic_detection} for proposed solution for microlearning which also considers consecutive session limits to minimize fatigue. In practical usage, designers also need to consider user safety, which is not captured in this diagram.}}
  \Description{A flowchart for automatic detection and utilization of opportunistic \browse{s} for mobile information acquisition.}	
  \label{fig:automatic_detection}	  
\end{figure}

\subsection{Design for information acquisition on OHMDs}

\subsubsection{Divided vs. Focused Attention} 
\begin{figure}[h]
  \centering
  \includegraphics[width=\linewidth]{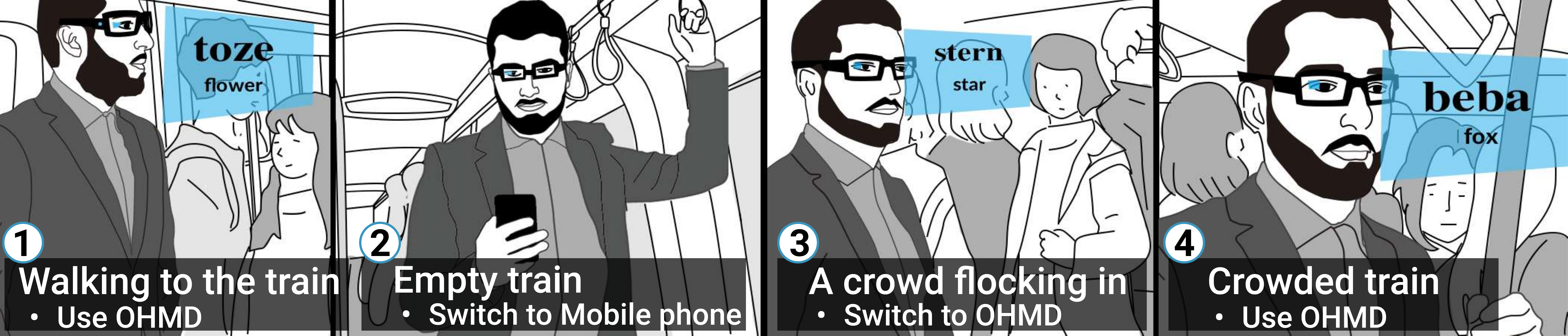}	  
  \caption{A potential scenario combining the use of OHMD and mobile phone to support commuters with microlearning}
  \Description{A potential combination of OHMDs with mobile phone to support microlearning in commute}
  \label{fig:design_implications_combined}	  
\end{figure}

Due to the dynamic change of \signal{s} on the move, we observed that users could engage in two information acquisition modes: \i{divided-attention mode} and \i{focused-attention mode}. Users enter the focused-attention mode when the frequency and intensity of the \signal{s} are low and the divided-attention mode when the frequency or intensity of \signal{s} is high. From our observations, OHMDs are suitable for the divided-attention mode as the see-through nature of the display screen allows users to use their peripheral vision to monitor \signal{s} and divide their attention between multiple tasks. On the other hand, mobile phones suit the focused-attention mode as users can focus on the screen without as many distractions from their surroundings. To strike a balance between the two modes, we propose two design solutions. The first design changes the OHMD foreground from transparent to opaque (i.e., words appear on an opaque overlay or ``light mode'' \cite{erickson_extended_2021}) when the focused-attention mode is triggered. This can improve users' concentration as it can block out the physical background and \signal{s}. The second design uses a combination system as shown in Fig~\ref{fig:design_implications_combined} to allow for flexible transitions between mobile phones and OHMD. However, we did not test the combination system and recommend further exploration of it. A balance needs to be struck between platform limitations and allowing users to maintain situational awareness as well as focus on display. Adjusting the opacity of OHMD content may help users focus on what is displayed but affect their level of situational awareness. Future investigations should undertake these design challenges.

\subsubsection{Peripheral notifications to minimize interruptions}
To render users more control of their OHMDs, information triggering applications can subtly notify users of any upcoming information acquisition sessions and allow them to ignore them if they wish. Moreover, to minimize the interruptions of such notifications, applications can either use nonvisual notifications or minimal UIs that reduce distractions from visual notifications. For example, NotifEye \cite{lucero_notifeye_2014} used a minimalistic playful user interface to present notifications on the user's focal vision while Luyten et al. \cite{luyten_hidden_2016} explored a visual language on peripheral vision. 

\subsubsection{Peripheral interactions for postponing or canceling}
Participants preferred using different interaction techniques to control microlearning sessions (e.g., continue, postpone, cancel) and learn content (e.g., word duration, appearing/disappearing) on OHMDs. The majority of participants did not prefer voice or touch input on the OHMD frame, all of which are standard OHMD interaction techniques, as they could be socially disturbing \cite{perficient_inc_mobile_2019} on public transport. Some participants preferred hands-free interactions, such as gaze and head movements, while a few suggested using the mobile phone as a controller. This indicates that there is a need for a set of socially acceptable and user-definable \cite{tung_user_defined_2015} OHMD interactions to cater to different information acquisition needs on the move. Furthermore, the interactions should not require any extra effort on users' part to initiate or stop since they should be able to divert their attention to their surrounding \signal{s} when necessary. Given that undesirable triggers can annoy users, we suggest using gaze behavior (e.g., blinking pattern) to cancel sessions since users can interact with the system without diverting their attention.

\section{Limitations}

We showed that our results apply to tech-savvy participants who are potential early adopters of OHMDs. Since device usage and motivation for information acquisition depend on users' technology acceptance and personal needs \cite{dornyei_motivation_1998, wixom2005theoretical}, we should generalize these results to other populations with care.

We also note that novelty effects might have affected the \glassprobe{}. Participants had only tested the OHMD during our training sessions before the actual experiment compared to a more extended history of mobile phone usage.

In addition, our observations were limited to a relatively small number of \added{(homogeneous)} participants \added{who had motivation to acquire information during commuting}, which may be insufficient for understanding the larger range of user behaviors. However, we believe that we have covered all prominent behaviors with our participant count, which achieved data saturation during the analysis.

\added{To mitigate the aforementioned limitations, large-scale longitudinal studies with different user groups (e.g., heterogeneous sample) and different OHMD prototypes are needed to identify the long-term effects and increase the generalizability of the results. Although our focus was not on micro-level eye movements (e.g., saccades, fixation \cite{kowler_eye_2011}), using advanced apparatus with eye-tracking (e.g., \cite{hansen_eye_2005}) can help to uncover the links between identified visual behaviors and micro-level eye movements and serve to build upon our proposed system (Fig~\ref{fig:automatic_detection}).} 


\deleted{We controlled the time gap between contextual inquiries in both studies and used longer observation/probe duration to minimize observer effects. There is, however, still the possibility of an observer effect. In addition, the controlled time gap between inquiries also limited us from exploring the participants' attitudes towards the frequency of consecutive microlearning sessions, though it provided participants with a realistic experience (given that users may not want to take on all information acquisition opportunities).}

\deleted{In this paper, we proposed visual behavior patterns based on video recordings, observations, and user opinions; nevertheless, a more accurate and finer visual attention model for each type of commuting activity can be developed using advanced eye-tracking technology (e.g., \cite{hansen_eye_2005}). Such studies will help unveil details at a granular level and serve to build upon our proposed system (Fig~\ref{fig:automatic_detection}).}

\section{Conclusions and Future Work}

Supporting ubiquitous information acquisition during on-the-move scenarios is challenging as users' surroundings dynamically change and can easily distract them. 
This research investigates common visual behaviors when users commute and differentiates between three types of visual behaviors: \glance{}, \observe{}, and \browse{}. We identified how they are affected by \signalforperception{s} and \signalforaction{s} in the environment and the resulting impact on users' receptivity to microlearning. The study also investigates the existing challenges of using two types of mobile platforms in such situations; the emerging OHMDs that provide enhanced information acquisition support on the move and existing mobile phones that offer more focused information acquisition with fewer \envinterrupts{}. Based on the limitations that exist in these platforms, we highlight the opportunities for technological advancement and better designs to support on-the-move information acquisition tasks. This research also suggests the potential coexistence of OHMDs with mobile phones to offer benefits beyond what either platform can individually provide. While microlearning is the domain of our investigation, we expect that our results apply to other forms of information acquisition aimed at improving users' productivity on the move.
Although this study mainly examines the visual behaviors of participants, we note that there are other essential factors such as social \cite{breen1985social}, affective \cite{knorzer_emotions_2016}, and environmental factors \cite{klatte2013does} that can influence users' receptivity to \i{information acquisition} and learning. These factors need to be considered as a whole to enhance information acquisition experiences effectively and help users become more receptive to such tasks. Therefore, we encourage longitudinal field studies for insight into how the aforementioned factors contribute to information \i{acquisition} and \i{retention} across platforms.

\section{\added{Data collection}}
\added{The data was collected on public transport just before the COVID-19 pandemic situation arose in Singapore (i.e., Sep. 2019 - Mar. 2020).}

\begin{acks}
This research/project is supported by the National Research Foundation, Singapore under its AI Singapore Programme (AI.SG Award No: AISG2-RP-2020-016). We thank Felicia Tan, Zhang Yue, and Mandalyn for their generous help with proofreading the drafts of paper and Zihan Yan for helping with a figure.
\end{acks}

\bibliographystyle{ACM-Reference-Format}
\bibliography{main}

\appendix

\section{Study 1: Shadowing}

\subsection{Video Coding}
\label{appendix:video_coding}

Please refer to Table~\ref{appendix:table:video_coding} for details.

\begin{table*}
\caption{Video coding for a participant (P1) up to 25 minutes. Note: Pattern/visual behavior was added later. Here \Intervene{} represents contextual inquiries that did not belong to natural behavior.}
\label{appendix:table:video_coding}
\begin{tabular}{p{0.1\textwidth} p{0.1\textwidth} p{0.1\textwidth} p{0.15\textwidth} p{0.35\textwidth} p{0.1\textwidth}  }
\hline
P1 & 9:25-11:05 & 11/6 & Bus/Walking & Home to school & \\ 
\hline
Time & Posture & Location & Familiarity/ Crowdedness & Visual behavior and primary task & Pattern \\ \hline
9:24:35 & Walking & Road & Familiar & looking around & \Browse{} \\
9:24:42 & Walking & Road & Less crowded & texting on phone & \Observe{} \\
9:24:53 & Walking & Road & & looking around (left and right)  & \Browse{} \\
9:26:07 & Walking & Road &  & reading on phone & \Observe{} \\
9:26:13 & Walking & Road &  & looking around & \Browse{} \\
9:26:23 & Walking & Road &  & check time on watch & \Glance{} \\
9:26:24 & Walking & Road &  & looking around & \Browse{} \\
9:26:31 & Walking & Road &  & texting, holding phone closer to eyes & \Observe{} \\
9:26:49 & Walking & Road &  & looking around & \Browse{} \\
9:27:43 & Walking & Road &  & texting on phone & \Observe{} \\
9:27:45 & Walking & Road &  & check for a bus & \Glance{} \\
9:27:46 & Walking & Road &  & texting on phone & \Observe{} \\
9:29:25 & Walking & Road &  & looking around & \Browse{} \\
9:30:02 & Walking & Road &  & engaging with phone & \Observe{} \\
9:30:07 & Walking & Road &  & check for a bus & \Glance{} \\
9:30:11 & Walking & Road &  & reading on phone & \Observe{} \\
9:31:01 & Walking & Road &  & looking around & \Browse{} \\ \hline
9:31:24 & Walking & Road &  & contextual inquiry & \Intervene{} \\ \hline
9:33:24 & Standing & Bus-stop & Less-crowded & engaging with mobile & \Observe{} \\
9:33:46 & Standing & Bus-stop &  & looking around & \Browse{} \\
9:33:49 & Standing & Bus-stop &  & check and notice the bus & \Glance{} \\
9:33:53 & Standing & Bus-stop &  & keep looking at the bus & \Observe{} \\
9:34:52 & Standing & Bus-stop &  & looking around & \Browse{} \\
9:35:35 & Standing & Bus-stop &  & glance at the coming bus & \Glance{} \\
9:35:37 & Walking & Bus-stop &  & looking around & \Browse{} \\
9:35:41 & Walking & Bus-stop &  & check the opening bus door & \Glance{} \\
9:35:42 & Walking & Bus-stop &  & keep looking at path/passengers, onboarding bus & \Observe{} \\
9:36:41 & Walking & Bus-stop &  & check for a seat & \Glance{} \\
9:36:45 & Walking & Bus-stop &  & keep looking at the surroundings to avoid bumping & \Observe{} \\ \hline
9:37:59 & Sitting & Bus & Many empty seats & looking around after sitting & \Browse{} \\
9:38:14 & Sitting & Bus &  & engaging with phone & \Observe{} \\
9:39:38 & Sitting & Bus &  & check the crowd & \Glance{} \\
9:39:41 & Sitting & Bus &  & doing nothing, looking ahead & \Browse{} \\
9:41:14 & Sitting & Bus &  & conversing with a neighboring passenger & \Observe{} \\
9:42:01 & Sitting & Bus &  & check the surrounded crowd & \Glance{} \\
9:42:03 & Sitting & Bus &  & doing nothing, looking ahead & \Browse{} \\
9:47:05 & Sitting & Bus &  & check the outside bus stop & \Glance{} \\
9:47:10 & Sitting & Bus &  & watching the outside scenery & \Observe{} \\
9:47:28 & Sitting & Bus &  & doing nothing, waiting for alight & \Browse{} \\
9:47:41 & Sitting & Bus &  & check the bus door & \Glance{} \\
9:47:45 & Sitting & Bus &  & looking around & \Browse{} \\ \hline
9:47:57 & Sitting & Bus &  & contextual inquiry & \Intervene{} \\ \hline
\end{tabular}
\end{table*}

\subsection{Data triangulation}
\label{appendix:study1:data_triangulation}

\begin{figure}[h]
  \includegraphics[width=\linewidth]{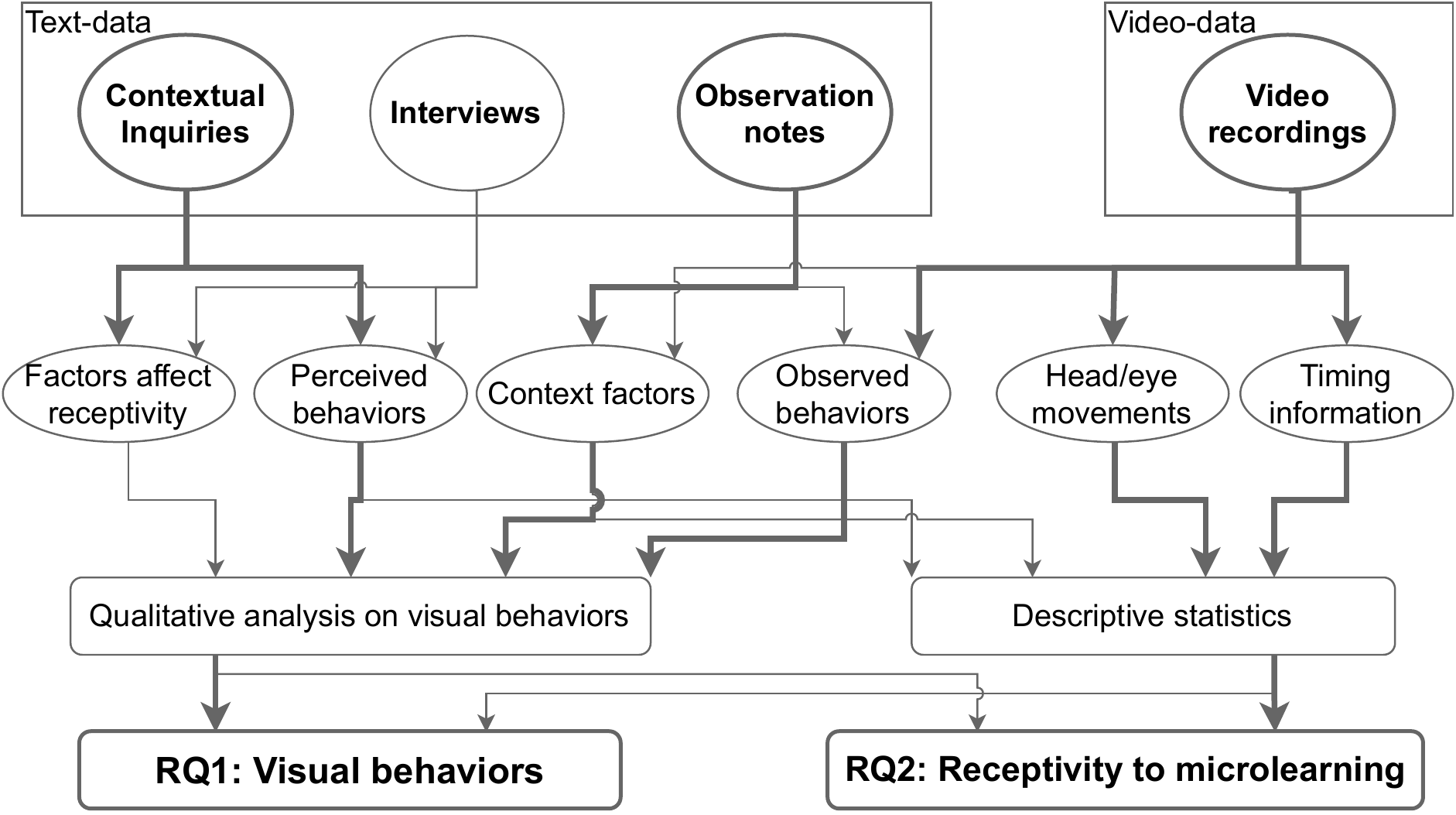}
  \caption{The triangulation of data sources towards research questions. The thickness of the lines/arrows represents the contribution of data to the analysis.} 
  \Description{The forms and triangulation of data towards research questions.}
  \label{fig:study1_triangulation}	  
\end{figure}

Fig~\ref{fig:study1_triangulation} shows the triangulation of four data sources; contextual inquiry notes, interview transcriptions, observation notes, and video recordings.

\subsection{Visual behavior distributions}
\label{appendix:visual_behavior_distributions}

The duration distributions of \observe{} and \browse{} were right-skewed and with log transformation, both approximated to normal distributions (\andersondarling{0.61}{0.64} for \observe{} and \andersondarling{1.12}{0.30} for \browse{}), while \glance{} did not (Table~\ref{table:study1_visual_behavior_distribution}). This finding aligns with literature as log-normal distributions commonly represent typical behaviors of humans, economics, biology, and so forth \cite{limpert2001log}.  

\Glance{} took the least amount of time, i.e., shortest duration (Table~\ref{table:study1_visual_behavior_distribution} and Fig~\ref{fig:behavior_frequency} (a) overall), though the frequency to duration (percentage) ratio was the highest. This implies that \glance{} has a supporting role for other behaviors given the higher frequency of occurrence.

\begin{table*}[h]
\caption{Distribution of visual behavior patterns; duration (in seconds; with order `$percentage, \: mean \: (sd) \: [min, max], \: median \: (IQR)$'), and frequency (`\i{percentage}, \i{patterns per hour}'). Note: Calculations of duration or frequency percentages were performed after removing the contextual inquiries since they were not part of the observed activities.}
\label{table:study1_visual_behavior_distribution}
\begin{tabular}{@{}lllll@{}}
\toprule
 Visual behavior pattern  & Distribution & \multicolumn{2}{l}{Duration} & Frequency  \\
\midrule
\Glance{}  & \multicolumn{1}{c}{-}  &   \visualduration{1.7}{1.2}{0}{4}{1.7} & , \medianiqr{1}{1}{3} & \visualfrequency{34.3}{33.2} \\
\Observe{} & $\sim$ log-normal & \visualduration{53.7}{83.4}{3}{695}{72.2} &  , \medianiqr{27}{12}{64} & \visualfrequency{48.3}{46.8} \\
\Browse{}  & $\sim$ log-normal & \visualduration{45.6}{48.4}{4}{302}{26.1} &  , \medianiqr{33}{13}{60} & \visualfrequency{20.6}{20.0} \\ 
\bottomrule
\end{tabular}
\end{table*}

\subsection{Statistics of visual behaviors}
\label{appendix:behavior_frequency}

Table~\ref{appendix:table:behavior_frequency} shows the associated data with Fig~\ref{fig:behavior_frequency}. 

\begin{table*}
\caption{Duration (in seconds; in order `$percentage, \: mean \: (sd) \: [min, max]$') and frequency (`\i{percentage}, \i{patterns per hour}') of visual behaviors of participants based on video analysis. All values are rounded to the nearest corresponding decimal place. $^a$ includes mobile language learning during the commute.}
\label{appendix:table:behavior_frequency}
\hspace*{-1cm} 
\begin{tabular}{p{0.11\textwidth} llllll}
\toprule
Behavior & \multicolumn{2}{l}{\Glance{}} & \multicolumn{2}{l}{\Observe{} $^a$} & \multicolumn{2}{l}{\Browse{}} \\ 
\hide{Participant} & Duration & Frequency & Duration & Frequency & Duration & Frequency \\ 
\midrule
\multicolumn{7}{c}{Overall} \\ \midrule
Overall & 
\visualduration{1.7}{1.2}{0}{4}{1.7} & \visualfrequency{34.3}{33.2} &
\visualduration{53.7}{83.4}{3}{695}{72.2} & \visualfrequency{48.3}{46.8} &
\visualduration{45.6}{48.4}{4}{302}{26.1} & \visualfrequency{20.6}{20.0} \\
P1 & 
\visualduration{1.0}{0.7}{0}{3}{1.7} & \visualfrequency{22.4}{22.8} &
\visualduration{47.6}{41.9}{3}{183}{50.1} & \visualfrequency{37.9}{38.6} &
\visualduration{45.8}{66.0}{4}{302}{48.2} & \visualfrequency{37.9}{38.6} \\
P6 & 
\visualduration{1.2}{0.9}{0}{4}{1.8} & \visualfrequency{55.2}{38.7} &
\visualduration{39.8}{42.9}{3}{220}{82.3} & \visualfrequency{74.5}{52.2} &
\visualduration{44.0}{31.8}{9}{117}{15.9} & \visualfrequency{13.0}{9.1} \\

\midrule
\multicolumn{7}{c}{Commuting Stage: Overall} \\ 
\midrule
Walking for commuting* & 
\visualduration{1.9}{1.4}{2}{4}{2.1} & \visualfrequency{39.0}{29.0} &
\visualduration{36.4}{37.7}{3}{206}{71.4} & \visualfrequency{70.6}{52.4} &
\visualduration{38.0}{32.4}{4}{117}{26.5} & \visualfrequency{25.1}{18.6} \\
Waiting for commuting & 
\visualduration{2.2}{1.2}{1}{4}{2.3} & \visualfrequency{36.9}{29.0} &
\visualduration{45.4}{48.3}{4}{199}{73.6} & \visualfrequency{58.4}{46.0} &
\visualduration{27.4}{20.1}{4}{74}{24.1} & \visualfrequency{31.8}{25.0} \\
On-vehicle & 
\visualduration{1.6}{1.2}{0}{4}{1.4} & \visualfrequency{32.0}{37.2} &
\visualduration{68.2}{109.9}{3}{695}{72.1} & \visualfrequency{38.1}{44.3} &
\visualduration{60.3}{61.4}{9}{302}{26.5} & \visualfrequency{15.8}{18.4} \\

\midrule
\multicolumn{7}{c}{Posture: Overall} \\ 
\midrule
Walking & 
\visualduration{2.1}{1.4}{0}{4}{1.7} & \visualfrequency{27.9}{24.3} &
\visualduration{41.6}{45.7}{3}{206}{76.1} & \visualfrequency{65.8}{57.2} &
\visualduration{37.6}{33.2}{4}{117}{22.2} & \visualfrequency{21.3}{18.5} \\
Standing & 
\visualduration{1.4}{1.1}{0}{4}{1.8} & \visualfrequency{45.2}{39.0} &
\visualduration{55.2}{105.8}{3}{695}{84.0} & \visualfrequency{54.8}{47.3} &
\visualduration{32.3}{20.0}{4}{70}{14.2} & \visualfrequency{15.9}{13.7} \\
Sitting & 
\visualduration{2.2}{1.3}{0}{4}{1.7} & \visualfrequency{26.8}{34.3} &
\visualduration{76.9}{79.6}{5}{287}{54.8} & \visualfrequency{25.7}{32.8} &
\visualduration{60.9}{65.7}{9}{302}{43.5} & \visualfrequency{25.7}{32.8} \\

\midrule
\multicolumn{7}{c}{P3 only} \\ 
\midrule
Overall & 
\visualduration{1.8}{1.2}{0}{4}{0.6} & \visualfrequency{12.4}{20.5} &
\visualduration{94.9}{170.5}{4}{695}{65.2} & \visualfrequency{24.7}{41.0} &
\visualduration{49.7}{37.7}{8}{158}{34.2} & \visualfrequency{24.7}{38.5} \\
\midrule
Road & 
\visualduration{2.3}{0.7}{2}{4}{2.5} & \visualfrequency{35.8}{28.6} &
\visualduration{36.0}{32.6}{5}{70}{53.7} & \visualfrequency{53.7}{42.9} &
\visualduration{44.0}{42.4}{14}{74}{43.8} & \visualfrequency{35.8}{28.6} \\
Station & 
\visualduration{1.0}{1.0}{0}{2}{0.4} & \visualfrequency{14.1}{9.0} &
\visualduration{26.1}{21.8}{4}{72}{71.8} & \visualfrequency{98.8}{63.6} &
\visualduration{50.5}{52.2}{15}{86}{27.8} & \visualfrequency{42.4}{27.3} \\
Metro & 
\visualduration{1.7}{1.4}{0}{4}{0.5} & \visualfrequency{10.3}{21.9} &
\visualduration{160.7}{226.2}{6}{695}{65.4} & \visualfrequency{14.7}{31.2} &
\visualduration{143.8}{55.0}{67}{205}{34.1} & \visualfrequency{22.0}{46.8} \\

\midrule
Walking & 
\visualduration{2.5}{0.7}{2}{3}{1.3} & \visualfrequency{18.8}{14.3} &
\visualduration{29.1}{20.1}{4}{70}{75.8} & \visualfrequency{93.7}{71.4} &
\visualduration{44.0}{42.4}{14}{74}{22.9} & \visualfrequency{18.8}{14.3} \\
Standing & 
\visualduration{1.0}{1.0}{0}{2}{0.1} & \visualfrequency{2.5}{9.1} &
\visualduration{294.5}{314.9}{16}{695}{83.1} & \visualfrequency{10.2}{36.4} &
\visualduration{39.7}{22.3}{8}{70}{16.8} & \visualfrequency{15.2}{54.5} \\
Sitting & 
\visualduration{1.7}{1.4}{0}{4}{1.1} & \visualfrequency{22.7}{28.0} &
\visualduration{71.5}{95.2}{6}{245}{38.6} & \visualfrequency{19.4}{24.0} &
\visualduration{150.7}{73.5}{67}{205}{60.3} & \visualfrequency{38.9}{48.0} \\

\bottomrule
\end{tabular}
\end{table*}

\subsection{Monitored Context Changes}
\label{appendix:study1_contexts}

\begin{itemize}
    \item Walking (after checking the road, while checking the road, after walking for a long time, after some filler activities such as phone engagements, walking fast/slow, walking in paths with turns and twists, walking in a crowd)
    \item Waiting for a vehicle (posture: standing/sitting, while checking the vehicle: a vehicle is/is not approaching, after checking the vehicle: correct/wrong vehicle, after checking the vehicle schedule: arriving/not arriving soon)
    \item On vehicle (posture: standing/sitting, same posture for some time: 1-2min/10min/30min/60min, after/while looking around/outside, doing nothing, after/during filling activities)
    \item Crowd changes (increase/decrease)
    \item Transitions (walking to/from the vehicle, walking to/from waiting, vehicle to/from waiting, vehicle to vehicle, few minutes before transitions, few minutes after transitions)
\end{itemize}

\subsection{Contextual Inquiry Topics}
\label{appendix:study1_contextul_inquiry}
\begin{itemize}
    \item Willingness to learn 3-6 words now (1-5 scale), reasons
    \item Factors affect the willingness, how they affect, why they affect
    \item Visual attention paid to the surroundings/context now (1-5 scale), reasons, whether it affects the willingness, if so how and why
\end{itemize}

\subsection{Interview Topics}
\label{appendix:study1_interview}

\subsubsection{Pre-study form and interview (for participants selection)}
\label{appendix:study1_interview:pre}
\begin{itemize}
    \item Demographics (age, gender, native language, education, employment)
    \item Commuting behaviors (mediums, activities that require commuting, duration/distance of commuting, and schedule)
    \item Second language learning (second language/s, reasons for learning, time allocated for learning, current fluency with the second language, apps using/used to support learning)
    \item Second language app usage (usage pattern, usage history, usage pattern during commuting, pain points of app usage in commuting, reasons for app usage/not usage)
    \item Factors affect language learning during commuting (factors, impact on learning, reasons)
\end{itemize}

\subsubsection{Post-study interview}
\label{appendix:study1_interview:post}
\begin{itemize}
    \item Reasons for visual behavior changes (explain the scenarios, if requires show video recording)
    \item Visual attention changes with crowdedness, path complexity, movements/motion, unfamiliarity, and posture
    \item Suitable/appropriate moments for microlearning during commuting, reasons/factors 
    \item Unsuitable/inappropriate moments for microlearning during commuting, reasons/factors
    \item Why not use moment (1,2...x) today for vocabulary learning (when they seem good for shadower), reasons
    \item Expectations for microlearning (minimum duration, number of words)
\end{itemize}



\section{Study 2: Technology Probe}

\subsection{Procedure}
\label{appendix:study2_procedure}
Before starting the probe, the experimenters provided a training session for the participants to familiarize themselves with the microlearning app on their respective platforms. Once each participant started commuting, two experimenters comprising the main experimenter and assistant experimenter followed them. After noting down context factors such as signal{s} (Appendix~\ref{appendix:study1_contexts}), the main experimenter would trigger a microlearning session when a \browse{} of more than 10-seconds was observed during commuting. A 10-second threshold was used to minimize recognition errors.
Then the main experimenter would conduct a contextual inquiry after the participant completes the microlearning session (see Appendix~\ref{appendix:study2_contextul_inquiry} for inquiry topics). The main experimenter kept the gap between the two triggers at a minimum of 5 minutes to reduce interruptions. On occasion, inquiries related to 2-3 consecutive sessions were conducted together (within 2-3 minutes). 
The assistant experimenter video recorded the whole study, mainly focusing on participant context and visual behaviors. 

At the end of the commute, the main experimenter carried out a 30-40 minute semi-structured interview with audio recording to assess the participants' experience with microlearning on their platform  (see Appendix~\ref{appendix:study2_interview} for interview topics). Whenever participants experienced difficulty recalling certain details, the experimenter replayed the specific instances from the video recordings to help them remember.


\subsection{Contextual Inquiry Topics}
\label{appendix:study2_contextul_inquiry}
\begin{itemize}
    \item Willingness to microlearning (1-5 scale), reasons
    \item Concentration on words (1-5 scale) and surroundings (1-5 scale), reasons, distractions from surroundings 
    \item Experience related devices after microlearning (comfortableness, issues, switching to learning, appearance/notifications), reasons
    \item Preference for presentation (appearing, disappearing, duration), reasons
    \item (For OHMD users) Experience/expectation with the mobile phone under similar situations, reasons
\end{itemize}

\subsection{Interview Topics}
\label{appendix:study2_interview}

\subsubsection{Pre-study form and interview (for participants selection)}
\label{appendix:study2_interview:pre}
\begin{itemize}
    \item Same questions used in \ref{appendix:study1_interview:pre}
    \item Mobile phone operating system
    \item Vision-related questions (vision corrections, usage of spectacles/contact lens)
\end{itemize}

\subsubsection{Post-study interview}
\label{appendix:study2_interview:post}
\begin{itemize}
    \item Experience of microlearning on the platform (overall, comfortableness of device usage, issues: social \& technical, automatic triggering, switching to learning), reasons
    \item Effects of situational factors and environmental interruptions on microlearning (situations easy/difficult to remember words, easy/difficult to concentrate on words), reasons
    \item Preference for presentation (appearing as a group, duration, number of words, learning vs. reviewing), reasons
    \item Expected interactions for the platform (controlling microlearning session, manual vs. automatic triggering)
    \item (For mobile phone users) Experience with notifications
    \item (For OHMD users) Experience/expectation/comparisons with the mobile phone under similar situations, reasons
\end{itemize}

\subsection{Utilizing \browse{s} for microlearning}
\label{sec:appendix:automatic_detection}

\begin{figure*}[h]
  \centering
  \includegraphics[width=\linewidth]{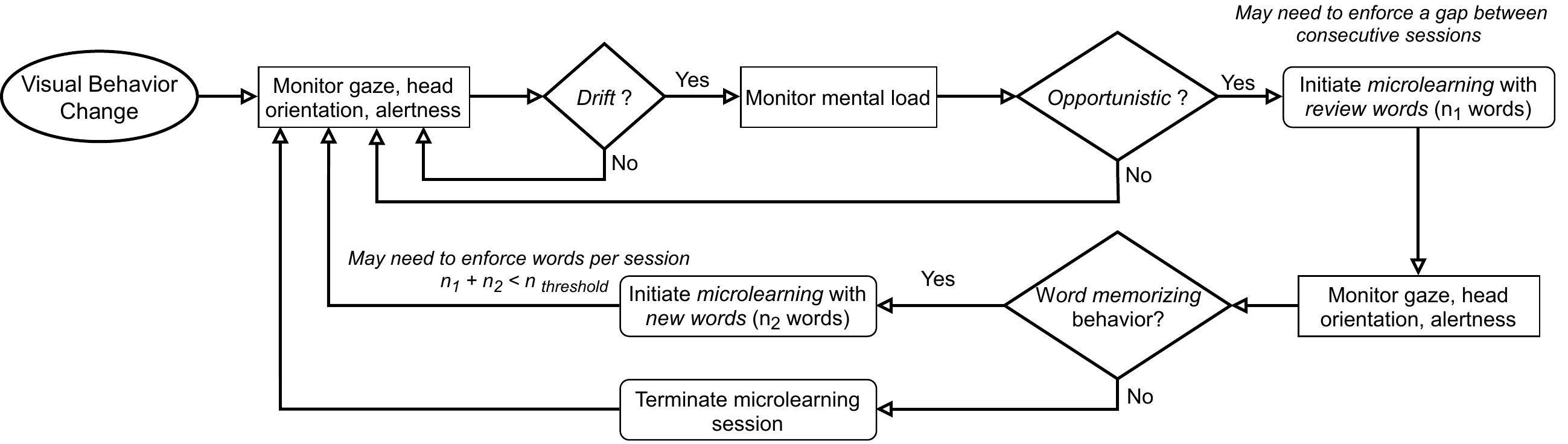}
  \caption{A flowchart for automatic detection and utilization of opportunistic \browse{s} for microlearning. 
  \i{`Opportunistic'} \browse{s} are determined by mental workload (sec~\ref{subsec:opportunistic_browse}) and sufficient duration (sec~\ref{subsec:behavior_and_receptivity}). Hence, before triggering the microlearning session, the duration of \browse{} behavior should exceed a customizable threshold. Additionally, conducting multiple sessions consecutively is potentially suboptimal since users may feel tired, affecting their ability to learn attentively. The maximum consecutive session count and the minimum time gap between sessions should be customized accordingly. Moreover, the number of words per session (learning or reviewing) should be customizable to cater to different user learning preferences. When the user requests to terminate the session in order to act upon dynamic \signal{s} (sec~\ref{subsec:automatic_manual}), the system should reset to the initial monitoring state. Practical implementation will also depend on the mobile platform since \i{`word memorizing behavior'} is influenced by the platform (sec~\ref{sec:interrupt_and_learn}). Further user studies are required to determine the subjective parameters, such as the time gap between sessions.}
  \Description{A flowchart for automatic detection and utilization of opportunistic \browse{s} for microlearning.}	
  \label{fig:appendix:automatic_detection}	  
\end{figure*}

Fig~\ref{fig:appendix:automatic_detection} shows our proposed solution for utilizing \browse{s} for microlearning.

\end{document}